\documentclass[twocolumn]{revtex4-1}
\usepackage[dvips]{graphicx}
\usepackage{latexsym}
\usepackage{graphicx}
\usepackage[export]{adjustbox}
\usepackage{float}
\usepackage[utf8]{inputenc}
\usepackage[english]{babel}
\usepackage[intlimits]{amsmath}
\usepackage{mathtools}
\usepackage{amssymb}
\usepackage{verbatim}
\usepackage{amsfonts}
\usepackage{hyperref}
\hypersetup{
	colorlinks=true,
	pdfborder={0 0 0},
}
\usepackage{enumitem}
\usepackage{color}
  \usepackage{verbatim}
     \newcommand{\Barnes}{\widetilde{G}}

  \definecolor{blue}{rgb}{0,0,1}
  \definecolor{green}{rgb}{0,.6,0}
  \definecolor{red}{rgb}{1,0,0}
  \definecolor{vio}{rgb}{1,0,1}
  \definecolor{uv}{rgb}{0.5,0,0.5}
  \definecolor{ama}{rgb}{0.3,0.3,0.3}
  \newcommand{\dif}{\mathrm{d}}
  \newcommand{\R}{\mathbb{R}}

   \newcommand{\im}{\mathbf{i}}
    
   \newcommand{\abs}[1]{\left\vert#1\right\vert}
   \newcommand{\set}[1]{\left\{#1\right\}}
   
   \newcommand{\vertex}{\mathcal{V}}
   \newcommand{\overlap}{\mathsf{q}}
   \newcommand{\defeq}{\stackrel{\text{def.}}{=}}
   \newcommand{\hoverlap}{\hat{\mathsf{q}}}

    \setlist[itemize]{leftmargin=*} 
\begin{document}

\title{Operator Product Expansion in Liouville Field Theory and Seiberg type transitions in log-correlated Random Energy Models}

\author{Xiangyu Cao}\affiliation{Department of Physics, University of California, Berkeley, Berkeley CA 94720, USA} \author{Pierre Le Doussal}\affiliation{CNRS-Laboratoire de Physique Théorique de l'\'Ecole Normale Supérieure, 24 rue Lhomond, 75231 Paris Cedex, France}\author{Alberto Rosso}\author{Raoul Santachiara}\affiliation{LPTMS, CNRS, Univ. Paris-Sud, Université Paris-Saclay, 91405 Orsay, France}
 \date{\today}
 \begin{abstract}
 We study transitions in log-correlated Random Energy Models (logREMs) that are related to the violation of a Seiberg bound in Liouville field theory (LFT): the binding transition and the termination point transition (a.k.a. pre-freezing). By means of LFT-logREM mapping, replica symmetry breaking and traveling-wave equation techniques, we unify both transitions in a two-parameter diagram, which describes the free energy large deviations of logREMs with a deterministic background log potential, or equivalently, the joint moments of the free energy and Gibbs measure in logREMs without background potential. Under the LFT-logREM mapping, the transitions correspond to the competition of discrete and continuous terms in a four-point correlation function. Our results provide a statistical interpretation of a peculiar non-locality of the operator product expansion in LFT. 
 The results are re-derived by a traveling-wave equation calculation, which shows that the features of LFT responsible for the transitions are reproduced in a simple model of diffusion with absorption. We examine also the problem by a replica symmetry breaking analysis. It complements the previous methods and reveals a rich  large deviation structure of the free energy of logREMs with a deterministic background log potential.  Many results are verified in the integrable circular logREM, by a replica-Coulomb gas integral approach. The related problem of common length (overlap) distribution is also considered. We provide a traveling-wave equation derivation of the LFT predictions announced in a precedent work.
 \end{abstract}

\maketitle
\section{Introduction}

Log-correlated random energy models (logREMs) form an extended class of simple disordered systems, which have applications in various problems of physics and mathematics, \textit{e.g.}, spin glass theory~\cite{derrida1980random,derrida1988polymers,derrida16kppfinitesize,derrida2017finite}, extremal properties of branching processes~\cite{krapivsky00kpp,arguin2012poisson,arguin2013extremal}, 2d XY model~\cite{carpenter98XY,carpentier00XYlong}, Anderson localization transitions~\cite{chamon1996localization,castillo97dirac,kogan96prelocalised}, random matrix and number theory~\cite{fyo12zeta,FyoKeat14,FyoSim15,ostrovsky2016riemann,arguin2017,arguin2017zeta}.
LogREMs share a host of glassy thermodynamic properties, which are reminiscent of, but considerably richer than, the uncorrelated Random Energy Model (REM)~\cite{derrida1980random}. In particular, all logREMs display a \textit{freezing} transition~\cite{derrida1988polymers,carpentier2001glass}: in the low-temperature glassy phase, the Boltzmann-Gibbs measure is dominated by a few \textit{atoms}, and the free energy becomes temperature independent. These features can be analytically understood by either replica symmetry breaking (RSB) or the velocity selection of traveling-wave equations of Fisher-Kolmogorov-Petrovsky-Piskunov (KPP) type.

A third intriguing approach is based on the connection between logREMs and Liouville Conformal Field Theory (LFT). This connection was proposed in Refs.~\cite{kogan96prelocalised,carpentier2001glass}, and is closely related to the probabilistic construction of LFT using the 2d Gaussian Free Field~\cite{david2016liouville,vargas_lectures,remy2017fyodorov}. Our previous work~\cite{cao16liouville,cao17thesis} revisited the connection and made it concrete. In particular we showed that LFT correlation functions provide the Gibbs measure statistics of a thermal particle in a 2d Gaussian Free Field plus an attractive deterministic logarithmic potential, a prototypical representative of the logREM class. Combined with freezing, this provided the first exact prediction of the minimum position of 2d Gaussian free field. Two types of transitions besides freezing were observed in that study: \texttt{i.} a binding transition~\cite{carpentier2001glass,fyodorov2009statistical,fyodorov2015moments} occurs when one of the charges, generating the deterministic potential, is sufficiently strong to trap the particle; \texttt{ii.} a termination point transition (also known as pre-freezing~\cite{fyodorov2009pre,wong17prefreezing}) related to the scaling of the $q$-th fractional moments of the Gibbs measure  (more precisely, the Gibbs probability weight). Indeed, their associated multi-fractal exponent $\tau_{q}$ saturates above a critical value $q \geq q_c$. This saturation, induced by the presence of atoms dominating the value of large moments, corresponds, in the LFT approach, to a competition between discrete and continuum terms in  the operator product expansion (OPE)~\cite{zamolodchikov1996conformal,teschner2001liouville,aleshkin2016construction}. This remarkable link has had multiple consequences, including: the prediction of universal log-corrections associated with the transition~\cite{cao16liouville,cao17thesis}, the extension to arbitrary temperature of recent results~\cite{derrida16kppfinitesize,derrida2017finite} on the overlap distribution of directed polymers on a Cayley tree, and the resolution of some standing puzzles concerning models such as the log-fractional Brownian motion~\cite{cao17fbm,fyodorov2015moments}.   

The purpose of this paper is to further explore the link between features of LFT and universal properties of the logREM class, by focusing on the statistics of the free energy of ``logREMs with one charge'', exemplified by the model of a thermal particle in a 2D Gaussian Free Field plus a deterministic logarithmic background potential generated by one attractive charge $a$. In particular, at a given temperature $1/\beta$, we determine the scaling of the $n$-th moment of the partition function, and find a rich diagram $(a,n)$ plane (Fig.~\ref{fig:phase}) consisting of four large deviation regimes; the scaling exponents undergo transitions (non-analyticity) on the regime boundaries. This model provides a common framework to study the binding and the termination point transitions.  We call both of them ``of Seiberg type'', for being related to the violation of the Seiberg bound in LFT~\cite{cao16liouville}. We argue that the regimes of this model are associated to different short-distance behaviors of the LFT correlation functions. Interestingly, one of them (the Bound regime defined below) is associated to LFT correlation functions whose short-distance singularity depends on the scaling dimension of distant fields. This non-locality is a consequence  of the LFT conformal bootstrap solution~\cite{zamolodchikov1996conformal,teschner2001liouville,aleshkin2016construction}, yet it has been little explored so far and requires an interpretation in terms of well-defined operator product expansions in a local conformal field theory. The application of this LFT property in logREMs is novel and generalizes significantly previous works.

We also compare the LFT approach to more conventional methods of disordered systems, i.e., replica symmetry breaking (RSB) and traveling-wave equations, with which we re-derive and complement the LFT predictions. It turns out that the traveling-wave approach is equivalent to the LFT one in their ability to calculate the universal quantities studied in this work, while RSB is different: it cannot calculate log-corrections to the leading behavior at the present stage, but can access the ``Log-Normal'' region~\cite{fyodorov2008statistical}, inaccessible from the other approaches.

The rest of the paper is organized as follows. Section~\ref{sec:synopsis} defines the models and the observables. Section~\ref{sec:LFT} presents the LFT approach, which allows to obtain a non-trivial part of the 2d diagram.  In Section~\ref{sec:rsb} we use the RSB method to obtain the complete diagram, confirming the LFT prediction. We also discuss extensively large deviation theory consequences of the results. Section~\ref{sec:kpp} confirms the LFT predictions using a traveling-wave equation approach and compare the three approaches. Section~\ref{sec:overlap} is devoted to the overlap distribution problem, and confirms the LFT prediction in Ref.~\cite{cao16liouville} by an independent, traveling-wave equation calculation.  The concluding Section~\ref{sec:discussion} is followed by a few appendices. In particular, Appendix~\ref{sec:morris} tests some predictions of the main text in an integrable logREM defined on the circle~\cite{fyodorov2015moments,ostrovsky2016gff,cao17fbm}. 


\section{The logREM class and its observables}\label{sec:synopsis}
The logREM class can be divided into two sub-categories, the \textit{Euclidean} ones and the \textit{hierarchical} ones. In both cases the disorder is given by a set of centered and correlated Gaussian random energy levels (or potential values), $\phi_j, j = 1, \dots, M$. The hierarchical logREMs is represented by the directed polymer on Cayley tree model, or the closely related Branching Brownian motion (BBM) model~\cite{derrida1988polymers}. Because of its close relation to the traveling-wave equation approach, we shall postpone their introduction to Section~\ref{sec:kpp}.

\subsection{Euclidean logREMs}
For Euclidean logREMs in $d$-dimension, $\phi_j$ is the discretization of a log-correlated field $\phi$ on a lattice $\set{\mathbf{r}_j}_{j=1}^M$ of lattice spacing $\epsilon=M^{-1/d}$ (so the large distance cut-off is of order-unity, $ \abs{\mathbf{r}_j} \leq R \sim O(1) $), so that the covariance decays logarithmically:
\begin{equation}
\overline{\phi_j} = 0\,,\, \overline{\phi_j \phi_k}  \sim \begin{cases}
- 2d \ln  \abs{\mathbf{r}_j - \mathbf{r}_k} &   \abs{\mathbf{r}_j - \mathbf{r}_k} \gg \epsilon \\
2 \ln M & j = k
\end{cases}  \,. \label{eq:logdecay}
\end{equation}
In particular, for $d = 2$, $\phi$ is known as the 2d Gaussian Free field (GFF), so 2d logREMs are closely related to its extreme values. 1d logREMs can be seen as obtained by restricting the 2d GFF on a 1d geometry (such as a circle~\cite{fyodorov2009statistical,cao15gff} or an interval~\cite{fyodorov2009statistical,fyodorov2015moments}); besides, 1d log-correlated fields (in the temporal domain) are also known as realizations of the $1/f$-noise. 
 
Given the random energy levels $\phi_j$, one defines the logREM by the partition function:
\begin{equation}
Z_0 \defeq \sum_{j=1}^M e^{-\beta \phi_j} \,.  \label{eq:Z}
\end{equation}
The above normalizations ensure that the freezing critical temperature is $\beta = \beta_c = 1$, and that the free energy has the following behavior~\cite{derrida1988polymers,carpentier2001glass}:
\begin{align}
& F_0\defeq -\beta \ln Z_0  = \begin{dcases}
- Q t + O(1) & \beta < 1 \\ -2 t + \frac32 \ln t +  O(1) & \beta > 1 \\
\end{dcases} \label{eq:freezing}  \\
&t \defeq \ln M \,,\, 
Q \defeq \beta + \beta^{-1}  \,,\, \beta < 1 \,; \,Q \defeq 2 \,, \beta > 1 \,. \label{eq:defQ} 
\end{align}
The notation $Q$ is inspired by LFT~\cite{cao16liouville}. The notation $t = \ln M$ is inspired by the Branching Brownian motion (see Section~\ref{sec:bbm}). In fact, $t$ is arguably a better measure (than the naive $M$) of the ``system size'' of logREMs, since the free energy is proportional to $t$; it is also known that the finite-size corrections in logREMs are of form $A_1/t + A_2 / t^2 + \dots$~\cite{cao16maxmin,derrida16kppfinitesize,mottishaw15REM}. We shall use both notations $t$ and $M$ in what follows. 

In eq. \eqref{eq:freezing}, $\frac32 \ln t$ is the universal log-correction characterizing the log-REM class~\cite{carpentier2001glass} (it becomes $\frac12 \ln t$ at $\beta = 1$), while ``$O(1)$'' denotes the order-unity fluctuating part of the free energy. Its distribution in the thermodynamic limit can be exacted calculated for BBM (see Section \ref{sec:kpp}) and for a few integrable logREMs~\cite{fyodorov2008statistical,fyodorov2009statistical,cao15gff}.

We are also interested in the random Gibbs probability weights 
\begin{equation}
p_{\beta,j} \defeq  \frac1 Z_0  e^{-\beta \phi_j} \,.
\end{equation}
Disorder-averaged moments and correlations of the logREM Gibbs probability weights were the key object in the LFT-logREM mapping~\cite{kogan96prelocalised,carpentier2001glass,cao16liouville}. In this work, we shall consider the joint moment of the Gibbs probability weights at a point and the partition function $\overline{p_{\beta,1}^q Z_0^n}$. 

 \subsection{LogREMs with charge}
 To study the binding transition, we define a logarithmic background potential (in short, {log potential} or background potential) for general logREMs, as follows:
\begin{equation}
U_j \defeq - a \, \overline{\phi_j \phi_1} \,,\, j = 1, \dots, M \,,\label{eq:defcharge}
\end{equation}
where $a > 0$ is called the \textit{charge}. We can check that this definition coincides with the same notion considered in Ref.~\cite{cao16liouville} for 2d log-REMs in a continuum setting. Indeed, eqs. \eqref{eq:logdecay} and \eqref{eq:defcharge} imply
$$ U_j =  - 2d a \ln \abs{z_j - z_1} =: U(z_j)  \,, $$ 
with $U(z) = -2d a\ln \abs{z-z_1}$. This expression, when $d=2$, reduces to the deterministic background potential with one charge inserted at $z_1$ defined in Ref.~\cite{cao16order} (for $d = 2$). We shall consider attractive potentials ($a > 0$), since only they can trigger a binding transition. We have picked $j=1$ to be the charge position by convention; in general, it does not correspond to a boundary of the lattice on which the logREM is defined.

With the log potential, we can define the logREM \textit{with one charge} by the following partition function summing over composite energy levels:
\begin{equation}
Z_a \defeq \sum_{j=1}^M e^{-\beta \varphi_j} \,,\, \varphi_j \defeq \phi_j + U_j \,. \label{eq:Za}
\end{equation}
As is known~\cite{carpentier2001glass,fyodorov2009statistical,cao16liouville}, the free energy $F_a \defeq -\beta^{-1} \ln Z_a$ undergoes the binding transition when $a = Q/2$: 
\begin{equation}
\overline{F_a} = \begin{cases}
-Qt + o(t) & a <  Q/2 \\ -2at + o(t) & a > Q/2 \,.
\end{cases} \label{eq:binding}
\end{equation}
That is, when $a > Q/2$ ($a< Q/2$), we have a Bound (Unbound) phase in which eq.~\eqref{eq:freezing} fails (holds, respectively). In Section~\ref{sec:rsb} we shall obtain the large deviation function of $F_a$ which puts the binding transition in a broader context.

\subsection{Girsanov transform}
The partition function moments of a logREM with charge is related to a joint moment of the logREM without charge by the Girsanov transform, also known as the complete-the-square trick (see Appendix~\ref{sec:girsanovapp}):  
\begin{equation}
\overline{ p_{\beta,1}^{a/\beta}   Z_0^{n + a/\beta}} = \overline{ e^{-a \phi_1}   Z_0^{n}}  =  M^{a^2} \, \overline{ Z_a^n } \,. \label{eq:girnasov}
\end{equation} 
The observables in this equation are the central object of this work, and allow to study both termination point and binding transitions:
\begin{itemize}
	\item[-] Setting $n = -a/\beta$, we obtain a fractional moment of the Gibbs probability weight $\overline{p_{\beta,1}^{a/\beta}}$, which undergoes a termination point transition when $a = Q / 2$; this transition has been studied by the LFT mapping in Refs.~\cite{cao16liouville,cao17thesis}. Via the Girsanov transform, the corresponding moment $\overline{Z_a^n} = \overline{Z_a^{-a/\beta}} = \overline{e^{a F_a}}$ describes positive large deviations of the free energy $F_a$. 
	\item[-] On the other hand, the binding transition concerns the \textit{typical} value of $F_a$, and is obtained by tuning $a$ across $Q/2$ while keeping $n$ infinitesimally close to $0$.
\end{itemize}

\section{The Liouville field theory approach}\label{sec:LFT}
We are interested in the asymptotic behavior of the joint distribution appearing in eq.~\eqref{eq:girnasov}, which will be shown to have the following general form:
\begin{equation}
\label{joint_distr}   \overline{ e^{-a \phi_1}   Z_0^{n}} \sim M^{-\Delta(a,n)}(\ln M)^{-\eta(a,n)}  \,.
\end{equation} 
Here and below, $\sim$ means equal up to an order-one factor as $M \to \infty$. In eq.~\eqref{joint_distr}, $\Delta(a,n)$ is the leading exponent and $\eta(a,n)$ the log-correction exponent. We will be working in the high-temperature $\beta < 1$ phase throughout this section.

\subsection{Summary of results}
\begin{figure}
	\includegraphics[width=.85\columnwidth]{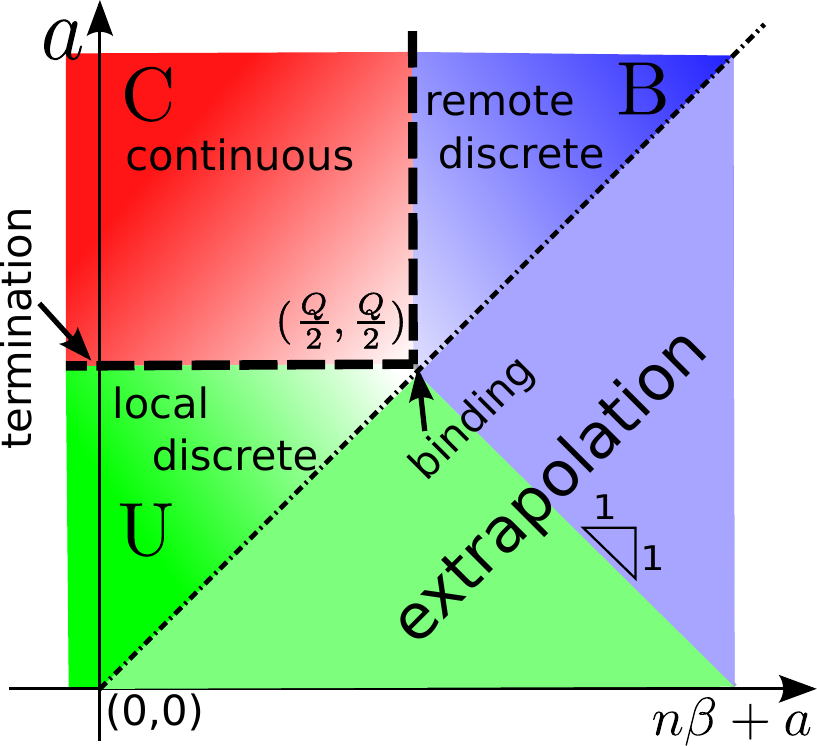}
	\caption{Diagram for the logREM observable $\overline{ p_{\beta,1}^{a/\beta}   Z_0^{n + a/\beta}} = \overline{ e^{-a \phi_1}   Z_0^{n}} $, eq.~\eqref{joint_distr}, obtained by the LFT approach (or by the traveling-wave equation approach, see section~\ref{sec:kpp}) . The parameter space is drawn with axes $n\beta + a$ and $a$. We indicate the three regimes (U, B, C), which are characterized by different dominating terms in the LFT correlation function, see Section~\ref{sec:nonlocal}. The locus of binding and termination point transition follow from the discussion below eq.~\eqref{joint_distr}. The LFT-logREM mapping is only valid in a half of the diagram with $n < 0$, i.e., above the dash-dotted diagonal. The rest of the diagram, drawn with pale colors, is a formal extrapolation, see eq.~\eqref{eq:LFTwrong}, and will be corrected by the RSB approach, see Fig.~\ref{fig:phase}.} \label{fig:phaseLFT}
\end{figure}

The LFT-logREM connection allows to determine the exponents $\Delta(a,n)$ and $\eta(a,n)$ in a sub-space of the $a,n$-parameter space delimited by $n < 0$. We find three regimes for the leading exponent:
\begin{align}
&\Delta(a,n) \stackrel{n<0}=  \nonumber \\
&\begin{dcases}
 -a^2 - Q n\beta & a < \frac{Q}{2}  \,, \\
 -(n\beta + a)^2    &   a + n \beta > Q / 2 \,, \\
 \frac{Q^2}{4} - Q (n\beta + a)  & a > \frac{Q}{2}  \,,\, a + n\beta < \frac{Q}{2} 
\end{dcases} \label{eq:Delta} 
\end{align}
Note that $\Delta(a,n)$ undergoes a second-order transition across a boundary between the regimes. We shall call the three regimes Unbound (U),  Bound (B), and Critical (C), see Fig.~\ref{fig:phaseLFT}. The Unbound (Bound) \textit{phase}, corresponding the {typical} free energy of the logREM with one charge, are described by the Unbound (Bound, respectively) regime near the line $n = 0$, respectively. The Critical regime (restricted to the $n\beta + a = 0$ line) was known as ``termination point''~\cite{cao16liouville,cao17thesis} or ``pre-freezing''~\cite{fyodorov2009pre,wong17prefreezing}, and is characterized by the log-corrections, as the exponent $\eta(a,n)$ is non-zero only in that regime and its boundaries:
\begin{align}
&\eta(a,n) \stackrel{n<0}= \begin{dcases} 
\frac{3}{2}  &  a > \frac{Q}{2} \,,\, a + n\beta < \frac{Q}{2}  \,, \\
\frac{1}{2}   &  a = \frac{Q}{2} \text{ or } a + n \beta = \frac{Q}{2}  \,, \\
0                 & \text{elsewhere}.
\end{dcases}\label{eq:eta}
\end{align}

Setting $ n = -a/\beta$ (i.e., along the vertical axis in Fig.~\ref{fig:phaseLFT}), we recover the results of Refs.~\cite{cao16liouville,cao17thesis} concerning fractional moments of $p_{\beta,1}$ (or equivalently, annealed averages of the inverse participation ratio). The leading order exponent in this special case
 \begin{equation} \Delta(a, -a/\beta)= \begin{cases}
a(Q-a) & a < Q/2  \\ Q^2/4 & a > Q/2
\end{cases}  \label{eq:lftdelta} \end{equation}
was at the origin of the logREM-LFT connection~\cite{kogan96prelocalised}. 

\subsection{Mapping to LFT}\label{sec:mappingLFT}
In order to obtain the above results, we shall consider a four-point correlation function of LFT with central charge $c=1+6Q^2$ defined on the sphere, which is equivalent to the complex plane plus infinity: $\mathbb{C}\cup \{\infty\}$. Using the same conventions as in \cite{cao16liouville},  we denote $\vertex_{a} (z)$ the LFT primary field with holomorphic and anti-holomorphic conformal dimensions $(\Delta_a,\Delta_a)$, with
\begin{equation}
\Delta_{a}=a(Q-a) \,. \label{eq:dimensionLFT}
\end{equation} 
In the standard LFT terminology, $\vertex_{a} (z)$ is called a vertex operator of charge $a$.  We will denote by $\left<\prod_{j}V_{a_j}(z_j)\right>_{\beta}$ an LFT multi-point correlation function. 

It is known~\cite{cao16liouville} that, in the high-temperature phase, $\beta<1$, one has the following relation between LFT correlation functions and observables of a logREM defined by a 2d GFF on the sphere: 
\begin{align}
 & \overline{\prod_{j}e^{a_j \phi_j}  
 Z_{0}^{n}} \sim M^{\sum_{j} a_j^2 + Q n \beta }\,
 \left<   \prod_{j}\vertex_{a_j} (z_j) \right>_{\beta} \label{eq:general} \\
 &\text{where }n = \frac{Q - \sum_{j} a_j}{\beta} < 0 \text{ and } a_j < Q/2 \text{ for all $j$.}   \nonumber
\end{align}
This statement can be shown using the Lagrangian representation of LFT, and a detailed exposition can be found in Section B of the supplemental material of \cite{cao16liouville}.  We note here that the factor $M^{\sum_{j} a_j^2 + Q n \beta }$ comes from the ultraviolet regularization of the exponential fields defined on the {\it discrete 2d GFF}:
\begin{align}
&\left(e^{a \phi_j}\right)_{\text{lattice}} \to M^{a^2}\left(e^{a \phi (z_j)}\right)_{\text{continuum}} \nonumber \\
& Z_0 = \sum_{j=1}^M e^{\beta \phi_j} \to M^{ Q \beta} \int \dif^2 z \, e^{\beta \phi (z)} \,,  \label{eq:uv}
\end{align}
where as in eq.~\eqref{eq:logdecay}, $\phi_j = \phi(z_j)$ is the discretization of the 2d GFF $\phi(z)$ on a lattice of points $(z_j)_{j=1}^M$ on the sphere. The inequalities in the last line of eq.~\eqref{eq:general} are known as the Seiberg bounds~\cite{seiberg1990notes}. 

To compute the exponents in eq.~\eqref{joint_distr}, we propose to consider the following LFT correlation function of two groups of $\ell$ and $m$ vertex operators:
\begin{align}
&\mathcal{K} \defeq \left< \prod_{j=1}^\ell \vertex_{a_j} (z_j)  \prod_{k=1}^m \vertex_{a_{\ell+k}}(w_k)    \right>_\beta \label{eq:lftob} 
\end{align}
where the charges satisfy 
\begin{equation} \sum_{j=1}^\ell a_j = a \,,\, \sum_{k=1}^m a_{\ell+k} = Q - n \beta - a \,. \label{eq:sum0} \end{equation} 
The points of the first group, $z_1, z_2, \dots, z_\ell$, are separated by lattice-space distance, $\abs{z_j  - z_i} = O(M^{-1/2})$, while all other pairs have order-unity separation. The number of operators in each group should be at least $2$, and sufficiently large to satisfy eq.~\eqref{eq:sum0} and the Seiberg bound $a_j < Q/2$. Further specifications of the parameters will turn out irrelevant. The idea is that, via the LFT-logREM mapping eq.~\eqref{eq:general}, the ``local group'' $\prod_{j=1}^\ell \vertex_{a_j} (z_j) $ produces the charge  $e^{-a\phi_1}$, and the ``remote group'' $\prod_{k=1}^m \vertex_{a_{\ell+k}}(w_k)  $ generates the correct power $n$ in $Z_0^n$. 

We now detail the above idea. For conciseness of the argument, let us restrict to the region $a <  Q$ and $a + n\beta > 0$. Then, the Seiberg bounds allow us to take $m = \ell = 2$, so that eq.~\eqref{eq:lftob} becomes a four-point function. To match the notations of Ref.~\cite{cao16liouville}, which we shall rely on in section~\ref{sec:nonlocal}, we re-order the indexes and consider the four-point function defined as follows:
\begin{align}
&\mathcal{K} = \left<   \vertex_{a_1} (0)  \vertex_{a_4} (z) \vertex_{a_2} (1) \vertex_{a_3} (\infty) \right>_{\beta} \,,\, \abs{z}^2 \sim M^{-1} \,,  \label{eq:K4} \\ 
&a_1 + a_4  = a \,,\, a_2 + a_3 = Q - n \beta - a \,.\label{eq:chargesums} 
\end{align}
Namely, $a_1, a_4$ form the local group and $a_2, a_3$ the remote group. The charge in eq.~\eqref{joint_distr} is now located near the origin of the complex plane.

Applying the general mapping~\eqref{eq:general} to $\mathcal{K}$ gives 
\begin{align}
&\overline{ e^{-a_1 \phi(0) - a_4 \phi(z)} Z_0^n  e^{-a_2 \phi(1) -a_3 \phi(\infty)} } M^{-a_2^2-a_3^2} \nonumber \\
& \sim M^{-\Delta_{a_1} - \Delta_{a_4} + Q(a + n \beta) }  \times \mathcal{K} \label{eq:K2}
\end{align}
where we used eq.~\eqref{eq:chargesums} and eq.~\eqref{eq:dimensionLFT} to rearrange the exponent of $M$.

 Then, we apply a Girsanov transform to the charges $a_2$ and $a_3$ (see Appendix~\ref{sec:girsanovapp} for more details), and obtain:
\begin{align}
&\overline{ e^{-a_1 \phi(0) - a_4 \phi(z)} \tilde{Z}_0^n }  \sim M^{-\Delta_{a_1} - \Delta_{a_4} + Q(a + n \beta) } \times  \mathcal{K}  \,. \label{eq:girsanov2}
\end{align}
Here $\tilde{Z}_0 = Z_0\vert_{\phi\to \tilde\phi}$, where $\tilde\phi$ differs from $\phi$ by a deterministic background potential [see eq.~\eqref{eq:U1infty}] which is smooth near $z= 0$, and has log-singularities at $z=1$ and $z=\infty$, with charges $a_2, a_3 < Q/2$. They satisfy the Seiberg bound, and are not strong enough to trigger a binding transition at either point [see eq.~\eqref{eq:binding}]. We thus expect that $ \tilde{Z}_0$ and $ Z_0$ lead to the same exponents $\Delta$ and $\eta$. This allows us to replace $\tilde{Z}_0^n$ by ${Z}_0^n$ in  eq.~\eqref{eq:girsanov2} and in what follows. Ultimately, we will justify this assumption self-consistently in the end of Appendix~\ref{sec:girsanovapp}. 

 Finally, since $\abs{z}= 1/\sqrt{M}$ is of lattice-size spacing, we can merge the two charges, i.e., perform the replacement $e^{-a_1 \phi(0) - a_4 \phi(z)} \leadsto e^{-a \phi_1}$ (recall $a = a_1 + a_2$) in the left hand side of eq.~\eqref{eq:girsanov2} without affecting the asymptotic behaviors  (this procedure is equivalent to the splitting method used in Ref.~\cite{cao16liouville}, Section 2.6.6). We thus obtain:
\begin{align}
\overline{ e^{-a  \phi_1 } {Z}_0^n }  \sim M^{-\Delta_{a_1} - \Delta_{a_4} + Q(a + n \beta) }  \times \mathcal{K} \label{eq:mapping} \,.
\end{align}
This is the main formula of this Section. It connects the exponents in eq.~(\ref{joint_distr}) to the asymptotic behavior of the four-point LFT four point function $\mathcal{K}$, which will be shown to be of the following form:
\begin{align} \label{eq:Kasym}
\mathcal{K} {\sim}  M^{\delta} \left(\ln M\right)^{-\eta} \;,\; \abs{z}^2 \sim 1/M \to 0 \,. 
\end{align}
Then eq.~\eqref{eq:mapping} implies
\begin{equation}
\label{DvsLFT}
\Delta(a,n)=-\delta + \Delta_{a_1} +\Delta_{a_4}-Q(n\beta+a) \,,\, \eta(a,n)= \eta \,.
\end{equation}

\subsection{Discrete terms and non-locality}\label{sec:nonlocal}
The exponents $\delta$ and $\eta$ depend on the value of $a_1,\dots, a_4$, and can be calculated by an analysis similar to Supplementary Material C.3 of Ref.~\cite{cao16liouville}, generalized in Appendix~\ref{sec:discrete} to take into account the \textit{remote discrete terms} produced by the group $2,3$. To explain the main idea, we first present the results for the leading exponent $\delta$:
\begin{align}
 &\delta = \Delta_{a_1} + \Delta_{a_4} - \Delta_{\alpha}  \,,\, \label{eq:deltaLFT}   \\
&  \alpha = \min (a_1 + a_4, a_2 + a_3, Q/2) \,. \label{eq:alphaLFT}
\end{align}
These results are obtained in Appendix~\ref{sec:discrete} by the conformal bootstrap solution of LFT. In this approach, ${\alpha}$ is known as the internal charge that dominates the four-point function. It is selected as having the smallest scaling dimension $\Delta_\alpha$ among the following candidates:
\begin{enumerate}
	\item  The dominant \textit{discrete term} produced by the \textit{local} group $1,4$, $\alpha = a_1 + a_4= a$, which exists \textit{if and only if} $a_1 + a_4 < Q/2$; 
	\item  The dominant \textit{discrete term} produced by the \textit{remote} group $2,3$, $\alpha = a_2 + a_3$, which exists \textit{if and only if} $a_2 + a_3 < Q/2$. 
	\item  The\textit{ continuous term} $\alpha = Q/2$, which is the dominant charge among the LFT spectrum $\alpha = Q/2 + \im P, P \in \R$.
\end{enumerate}
Combining eqs.~\eqref{DvsLFT}, \eqref{eq:dimensionLFT} and eq.~\eqref{eq:deltaLFT}, we obtain the leading exponent in eq.~\eqref{joint_distr}:
\begin{equation} \Delta(a,n) = \alpha (Q-\alpha) - Q (a+ n\beta) \,. \label{eq:Delta1}\end{equation}
The three regimes in eq.~\eqref{eq:Delta} can be characterized by the choice of dominant $\alpha$ [see also eq.~\eqref{eq:chargesums}]:
\begin{enumerate}
\item Unbound: when $a_1 + a_4 = a < Q/2$, the local discrete term dominates: $\alpha = a$.
\item Bound: when $a_2 + a_3 < Q/2\Leftrightarrow  n \beta + a > Q/2$, the remote discrete term dominates: $\alpha = Q-n\beta -a$.
\item Critical: when  $a>Q/2$ and $n \beta + a < Q/2$, no discrete term is present, so the continuous term dominates: $\alpha = Q/2$.
\end{enumerate}
One can readily check that eq.~\eqref{eq:Delta1}, supplemented by the above values of $\alpha$, agrees with eq.~\eqref{eq:Delta}. To obtain the log-corrections, we recall~\cite{ribault2015liouville,cao16liouville} [see also Appendix~\ref{sec:discrete}, below eq.~\eqref{eq:Cont}] that it is absent in the LFT correlation function~\eqref{eq:Kasym} ($\eta = 0$) when a discrete term dominates, and is present with exponent $\eta = 3/2$ when dominated by the continuous term; finally, $\eta = 1/2$ in the marginal case corresponding to regime boundaries. Eq.~\eqref{eq:eta} follows readily from these facts and the regime characterizations above. 

For the sake of comparison with the RSB approach (in Section~\ref{sec:rsb}), we note that the above analysis can be formally extended to the region $n \geq 0$, by ignoring the bound $n<0$ in eq.~\eqref{eq:general}. Then, both discrete terms will be present and compete with each other, resulting in the following formal extrapolation of eq.~\eqref{eq:Delta}:
\begin{align}
&\textbf{Formally, } \Delta(a,n) \stackrel{n\geq0}= \nonumber \\
& \begin{cases}
a(Q-a) - Q (n\beta + a) & 2a + n \beta > Q \\
-(n\beta + a)^2    &   2a + n \beta < Q  
\end{cases} \label{eq:LFTwrong}
\end{align}
The new boundary $2a + n \beta = Q , n > 0$ separating the U and B regimes is generated by the competition between the two discrete terms. 

To close, we remark the ``non-locality'' of the asymptotic behavior of the four-point correlation function above. Since only the local fields $1$ and $4$ are approaching each other, one would be tempted to expect that the asymptotic behavior  depends only on $a_1$ and $a_4$, but not on the remaining fields in the correlation function. However, in LFT, and more generally, in conformal field theories, such as the minimal models~\cite{ribault2014conformal}, this is not necessarily the case. As we illustrate in Appendix~\ref{sec:discrete}, the conformal bootstrap solution of LFT implies that when a remote discrete term is present, the OPE depends explicitly on the remote charges.  This property is crucial for describing correctly the Bound regime. To our knowledge, such an application to disordered statistical mechanics was not noticed before (the importance of discrete terms for the consistency of LFT is well understood, see e.g. Ref.~\cite{teschner2001liouville}). In the following we shall corroborate the LFT predictions by two independent methods.

\section{Replica symmetry breaking}\label{sec:rsb}
In this Section, we calculate the leading exponent of eq.~\eqref{eq:girnasov} in the full parameter space, by one-step replica symmetry breaking. This method has the advantage of covering the full parameter space, which allows to explore the Log-normal regime, invisible to the other approaches.  In Section~\ref{sec:rsb1}  we go through the RSB calculation and obtain the complete diagram. The two remaining subsections discuss the physical/probabilistic meaning of the results in terms of large deviation theory. 

\subsection{Determination of the complete diagram}\label{sec:rsb1}
\subsubsection{One-step replica symmetry breaking}\label{sec:rsb11}
Replica symmetry breaking can be applied to logREMs, Euclidean as well as hierarchal, because all logREMs enjoy an asymptotic ultra-metricity property, see Ref.~\cite{cao17thesis} (section 2.3.2) for explanation. It is useful to introduce the notion of \textit{overlap} $\overlap_{ij}$ between two sites of a logREM, defined as:
\begin{equation}
\overlap_{ij} \defeq \frac{1}{2 t} \, \overline{\phi_i \phi_j} \,, \, t \defeq \ln M \,. \label{eq:overlap}
\end{equation}
By eq.~\eqref{eq:logdecay}, when the two sites $\mathbf{r}_i$ and $\mathbf{r}_j$ are separated by a distance of lattice spacing order, $\overlap_{ij} \to 1$; when they are far away, and the covariance between $\phi_i$ and $\phi_j$ is of order unity, $\overlap_{ij} = 0$; in general, $\overlap_{ij} \in [0, 1]$. The terminology originates from hierarchal logREMs, see eq.~\eqref{eq:commonlength} below. 

The starting point of the RSB method is the following exact formula for eq.~\eqref{eq:girnasov} when $n = 1, 2, 3,\dots$, which can be obtained by the Wick theorem:
\begin{align}
&\overline{ e^{-a \phi_1}   Z_0^{n}} =  \label{eq:wick} \\
&\sum_{(j_\mu)_{\mu=1}^n} \exp\left[\frac{\beta^2}2 \sum_{\mu, \nu=1}^n \overline{\phi_{j_\mu} \phi_{j_{\nu}}} + \frac{a^2}{2} 
\overline{\phi_1^2} + \beta a \sum_{\mu=1}^n \overline{\phi_1 \phi_{j_\mu}}  \right] \nonumber
\end{align}
We need to estimate the $M \to \infty$ asymptotics of this replica sum, when $n$ is considered to be analytically continued to a real value.  The techniques involved are quite peculiar and heuristic, yet relatively well-known in the spin-glass literature, see e.g. Refs.~\cite{dotsenko1995introduction,bouchaud1997universality}.  For logREMs without charge, the replica approach was developed in Refs.~\cite{carpentier2001glass,fyodorov2010freezing,cao16order} (see Ref.~\cite{cao17thesis}, Section 2.3 for a more pedantic introduction).

\begin{figure}
\begin{center}
	\includegraphics[width=.85\columnwidth]{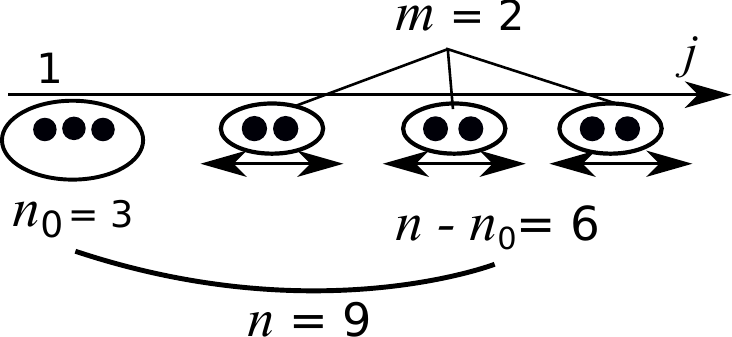}
\end{center}
	\caption{Illustration of a dominating configuration according to the one-step RSB Ansatz. The horizontal axis represents the space on which is defined the logREM potential. The circles represent the position $j_\mu$ of the $n$ replicas. The $n_0$ replicas in the circle are attached to the charge at $j = 1$. The other $(n - n_0)$ replicas form groups of size $m$, whose positions are not fixed. Section~\ref{sec:Op1} discusses the RSB of the $(n-n_0)$ replicas away from the charge (optimization of $m$); Section~\ref{sec:Op2} discusses the optimization of $n_0$.}\label{fig:rsb}
\end{figure}

For our situation with one charge, we shall adopt and extend the \textit{one-step} RSB Ansatz proposed in Ref.~\cite{fyodorov2009pre}. According to it, the dominant terms in eq.~\eqref{eq:wick} have the positions of the $n$ replicas $j_1, \dots, j_n$ organized in the following way (an illustration is provided in Fig.~\ref{fig:rsb}.:  $n_0$ of them are attached to the charge $a$, and have overlap $1$ with the site $1$; the rest of replicas are not affected by the charge, and form groups of size $m$ that are free to move in the system. More precisely, replicas of the same group have mutual overlap $\overlap=1$ and different groups have mutual overlap $0$, and all groups have overlap $0$ with site $1$. The attribute ``one-step'' refers to the fact that the overlaps assume only two values (so the permutation symmetry between the replicas is spontaneously broken ``once''): $0$ and $1$. 

Under the above one-step RSB Ansatz, eq.~\eqref{eq:girnasov} is evaluated to have the following leading behavior:
\begin{align}
 \left.\overline{ e^{-a \phi_1}   Z_0^{n}}\right\vert_{m,n_0} &\approx M^{- \Delta(a,n \vert m,n_0)} \,,\,  \label{eq:hmndef} \\
 \Delta(a,n\vert m,n_0) &= -(a + n_0 \beta )^2  \nonumber \\ 
-&  \frac{n-n_0}{m} \left(1 + m^2\beta^2 \right)  \,,\,  \label{eq:hmn} 
\end{align} 
In the right hand side of eq.~\eqref{eq:hmn}, $(a + n_0 \beta )^2$ corresponds to the Wick contractions between the $n_0$ replicas near site $1$ and the charge, and the remaining term corresponds to the other $(n-n_0)/m$ groups of replicas. In the bracket, $1$ is an entropic term (the constraint that different groups cannot have overlap $>0$ does not affect the leading behavior); $m^2\beta^2$ comes from the Wick contractions between replicas of the same group. The Wick contractions between replicas from different groups do not affect the leading exponent (because of their vanishing overlap) and are ignored in eq.~\eqref{eq:hmn}.
The notation $\Delta(a,n \vert m, n_0)$ is chosen in relation to $\Delta(a,n)$ in eq.~\eqref{joint_distr}. They both denote the leading exponent, yet $\Delta(a,n \vert m, n_0)$ depends in addition on the \textit{variational parameters} $m$ and $n_0$, that we need to optimize, according to the non-rigorous optimization rules of RSB that we discuss below. The result of the optimization will be (the RSB prediction of) $\Delta(a,n)$:
 \begin{equation}
 \Delta(a,n)\vert_{\text{RSB}} =   \Delta(a,n \vert m, n_0)\vert_{m, n_0 \text{ optimized}} \,. \label{eq:DeltaRSB}
 \end{equation}

\subsubsection{Optimization of $m$: freezing transition and Log-Normal regime}\label{sec:Op1}
We first review the rules for the optimization of the second term of eq.~\eqref{eq:hmn} with respect to $m$, with $n-n_0$ fixed. These rules concern the replicas not attached to the charge. Equivalently, one may consider logREMs without charge, corresponding to the special case $a = 0, n_0 = 0$. The RSB rules are known to be the followings:
\begin{enumerate}
\item $m$ is between $n-n_0$ and $1$: $m \in [n-n_0, 1]$ if $n-n_0 < 1$ and $m \in [1, n-n_0]$ if $n-n_0> 1$.
\item $\Delta(a,n\vert m, n_0)$  is \textit{maximized} with respect to $m$ when $n-n_0 < 1$ and minimized otherwise. 
\end{enumerate} 
We now apply these rules to eq. \eqref{eq:hmn} with fixed $n-n_0$. As a function of $m$, eq.~\eqref{eq:hmn} has a unique maximum at $m = 1/\beta$. According to the above rules, this is the actual optimal $m$ if and only if $n-n_0 < 1/\beta < 1$. In all other cases, the optimal $m$ is one of the boundaries $n-n_0$ or $1$. A thorough case study leads to the results summarized in Table~\ref{eq:bulk}; in particular, we find the solutions are smooth across $n-n_0=1$ despite the change of the rules.
\begin{table}
\begin{tabular}{|c|c|c|c|}
\hline
$\beta$ & $n-n_0$  & $m$ &  $-\Delta(a,n\vert m, n_0) - (a + n_0 \beta )^2$   \\
\hline 
$< 1$ & $<\beta^{-2}$ &  $1$ &  $ (n-n_0) (1+ \beta^2) $   \\
$ < 1$ & $>\beta^{-2}$ &  $(n-n_0)$ &  $ 1 + ((n-n_0)\beta)^2$   \\
$ > 1$ & $< \beta^{-1}$ & $\beta^{-1}$ &  $2 (n-n_0)\beta$  \\
$> 1$ & $> \beta^{-1}$ & $(n-n_0)$ & $1 + ((n-n_0)\beta)^2$ \\
\hline
\end{tabular}
\caption{One-step replica symmetry breaking solutions for (Euclidean) logREMs without charge. The first case is the replica-symmetric solution. The third case is the replica symmetry breaking phase. The second and fourth cases correspond to the Log-Normal solution in $\beta<1$ and $\beta>1$ phases respectively.} \label{eq:bulk}
\end{table}

The results of Table~\ref{eq:bulk}, specialized to the case $a = 0$ and $n_0 = 0, n-n_0 = n$, correspond to known facts of logREMs without charge, which we review in the rest of this section. The typical free energy of the bulk, corresponding to $n \sim 0$, is governed by the {replica symmetric} solution $m=1$ and the replica symmetry breaking solution $m=1/\beta$ in the high-temperature ($\beta <1$) and the frozen ($\beta > 1$) phase, respectively. On the other hand, the negative large deviation of the free energy, corresponding to $n \gg 0$, is governed by the \textit{Log-Normal} solution~\cite{fyodorov2008statistical,cao17thesis} $m = n$. We call it Log-Normal because it corresponds to a Gaussian far tail of the free energy, and thus a log-Normal tail of the distribution of the partition function~\cite{fyodorov2008statistical}. In the high-temperature phase, the transition at $n \beta = 1/\beta$  is due to an exponential left tail of the distribution of the free energy centered around its typical value: $f_0 = F_0 + Qt, t := \ln M$ [see eq.~\eqref{eq:freezing}]:
 $$P(f_0) \sim e^{f_0/\beta} \,,\, f_0 \to -\infty \,. $$ 
 This tail is at the origin of the divergence of $\overline{Z_0^{n}} \propto \overline{\exp( - n \beta f_0 )}$ in continuum replica calculations~\cite{fyodorov2008statistical}. In the frozen phase $\beta > 1$, the tail acquires a log correction and its exponent becomes independent of $\beta$~\cite{carpentier2001glass,fyodorov2008statistical}: $$P\left(f_0 = F_0 + Qt - \frac32 \ln t \right) \sim \abs{f_0} e^{f_0} \,,\, f_0 \to -\infty \,.$$ 
 Therefore the transition to Log-Normal regime happens at $n \beta = 1$. As we shall see below in Section~\ref{sec:LDF1} and Appendix~\ref{sec:morris}, exponential tails are a characteristic signature of {first-order} transitions between different regimes describing the negative large deviations of the free energy. We will also generalize the above results to the case of  logREMs with one charge.

\subsubsection{Optimization of $n_0$}\label{sec:Op2}
\begin{figure}
	\includegraphics[width=.85\columnwidth]{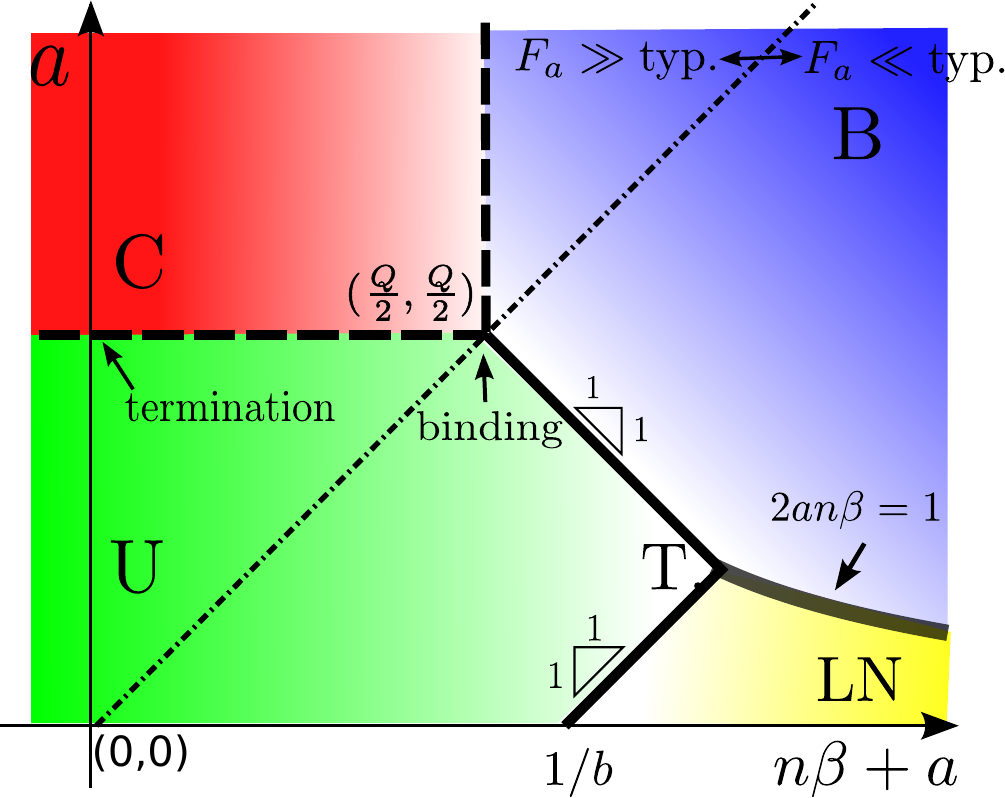}
	\caption{Complete large deviation diagram of the logREM with a charge for any fixed temperature $1/\beta$, obtained by computing the observable $\overline{ e^{-a \phi_1}  Z_0^{n}} = M^{a^2} \overline{Z_a^n}$ [see eqs.~\eqref{eq:girnasov} and \eqref{eq:Za}] by the RSB approach. The axes are the same as in Fig.~\ref{fig:phaseLFT}. The four regimes: Unbound (U), Critical (C) and Bound (B), and Log-Normal (LN), drawn in different colours, are separated by 1st (solid) and second order (dashed)  transitions. The diagonal $n = 0$ describes the typical fluctuation of the free energy $F_a$ of the logREM with one charge. The binding transition happens when crossing the tri-critical point $(Q/2, Q/2)$ along the diagonal (the full large deviation theory of $F_a$ is discussed in Sec.~\ref{sec:LDF1}). The termination point transition of the Gibbs probability weight happens when crossing U-C boundary along the $a$-axis. The region $n > 0$ is not accessible in the LFT approach, whose formal prediction misses the Log-Normal regime, see Fig.~\ref{fig:phaseLFT}. We denoted $b = \min(1, \beta)$ and $Q = b + b^{-1}$ in agreement with eq.~\eqref{eq:defQ}. The triple (T) point at which join the U, LN and B regime boundaries has coordinates $(n\beta + a = 1/b+ b/2, a = b/2)$.} \label{fig:phase}
\end{figure}

Now we come to the optimization of $n_0$. Since it concerns the interaction with the log-potential and not amongst the replicas themselves, the rules are different from those for $m$:
\begin{enumerate}
\item $n_0$ is between $n$ and $0$: $n_0 \in [0, n]$ if $n > 0$ and $n_0 \in [n, 0]$ if $n < 0$.
\item $\Delta(a,n\vert m, n_0)$ is minimized (maximized) with respect to $n_0$ when $n > 0$ ($n<0$, respectively).
\end{enumerate}
We apply these rules on top of the previous results in Table~\ref{eq:bulk}. The case analyses are completely elementary but cumbersome to enumerate in detail. The key point is to consider separately $n < 0$ and $n > 0$, and observe that the Log-Normal solution ($m = n - n_0$) can only occur in the $n > 0$ case; otherwise, $m = 1/b$, with $b = \min(1,\beta)$ independently of $n_0$. Going through all cases, we obtain the complete diagram depicted in Fig.~\ref{fig:phase}; the optimal parameters and the leading behavior of eq.~\eqref{eq:girnasov} are provided in Table~\ref{tab:res}. 

We now compare the above results with the LFT ones. The leading exponent obtained by RSB agrees with the LFT prediction~\eqref{eq:Delta} in the whole region $n < 0$ allowed by the LFT-logREM mapping. This is remarkable given the difference of the two methods: no intermediate steps can be compared. We can push the comparison to the region $n \geq 0$, where the LFT approach loses its validity in principle. Even then, the {formal} LFT prediction, eq.~\eqref{eq:LFTwrong}, still agrees with the RSB result outside the Log-Normal regime, which is neglected by LFT and can be only accessed within the RSB approach.

\begin{table*}
	\begin{tabular}{|c|l|c|c|c|c|}
		\hline
		Regime & Locus  & $n_0 \beta$ & $m \beta$  & $\Delta(a,n)$ & $ \tau(s) \defeq \lim_{t\to\infty}\left( \frac1{\ln M} \ln  \overline{e^{s F_a}} \right) $   \\ \hline
		Unbound  &  $a < \frac{Q}{2} -(n\beta)_+, n \beta < \frac1{b}$    & $0$ & $b$ & $ -a^2 - Q n\beta $ &  $ - Q s$\\ \hline 
		Critical   &  $0 < a-\frac{Q}2 < -n\beta$  & $\frac{Q}2-a$ & $b$  & $\frac{Q^2}{4} - Q (n\beta + a) $  & $-Q s - (\frac{Q}2-a)^2 $ \\ \hline
		Bound &    $a > \max(\frac{Q}2 - n \beta,\frac{Q}2 - \frac{n \beta}{2}, \frac{1}{2n\beta} )$ & $n \beta$ & $b$  & $-(n\beta + a)^2$ & $s^2-2as$ \\ \hline
		Log-Normal  &  $\frac1{b} < n \beta < \frac1{2a}  $     & $0$ &  $n \beta$ & $-a^2 - n^2\beta^2 - 1$ &  $ s^2 + 1 $\\ \hline
	\end{tabular}
	\caption{Summary of the results of RSB analysis. See Fig.~\ref{fig:phase} for an illustration of the locus of each regime. The third and fourth columns give the optimal value of the variational parameters $m,n_0$. The fifth column gives the RSB prediction of the leading exponent, eq.~\eqref{eq:DeltaRSB}. For compactness of the expressions, we denoted $b \defeq \min(\beta,1)$, so $Q = b + b^{-1}$, as in eq.~\eqref{eq:defQ}, and $(x)_+ = \max(x,0)$. The last column is obtained by applying $\tau(s) =- \Delta(a, -s/\beta) - a^2$, eq.~\eqref{eq:tauDelta}.} \label{tab:res}
\end{table*}

Another advantage of the RSB approach is that its solution provides more direct physical insights (than the LFT approach). Indeed, the nature of different regimes can be understood by examining the optimized value of the variational parameters. Let us focus on $n_0$, which we recall is the number of replicas that are attached to the charge [see above eq.~\eqref{eq:hmndef}]:
\begin{itemize}
	\item[-]  In the Unbound regime, $n_0 = 0$: no replica is attached to the charge, so the thermal particle is unbound. In particular, in the Unbound regime of the high temperature $\beta<1$ phase, the replica symmetry is preserved everywhere. 
	\item[-] In the Bound regime, $n_0 = n$: all replicas are attached to the charge, so the thermal particle is trapped at the bottom of the log-potential.
	\item[-] The Critical regime is the only one in which $ n_0 $ attains its stationary point (i.e., $\partial_{n_0} \Delta = 0$)  
	\begin{equation} \left( n_{0} \beta \right)_{\text{Critical}} = Q/2 - a \,, \label{eq:RSBsaddle} \end{equation}
	which is in the interior of the allowed interval: $ n < n_0 < 0$. In this respect this regime describes a ``Critical'' state between Unbound and Bound regimes (more precise statistical interpretations will be provided in Sections~\ref{sec:LDF1} and \ref{sec:LDF2}). 
\end{itemize} 
The presence of the Critical regime adjacent to the binding transition point ($a = Q/2, n = 0$, see Fig.~\ref{fig:phase2}) makes the latter the most intricate spot of the whole 2d diagram: it is a tri-critical point. To our knowledge, such a non-trivial nature of the binding transition is revealed here for the first time.   

\subsection{Large deviations of $F_a$} \label{sec:LDF1}
\begin{figure}
	\includegraphics[width=.8\columnwidth]{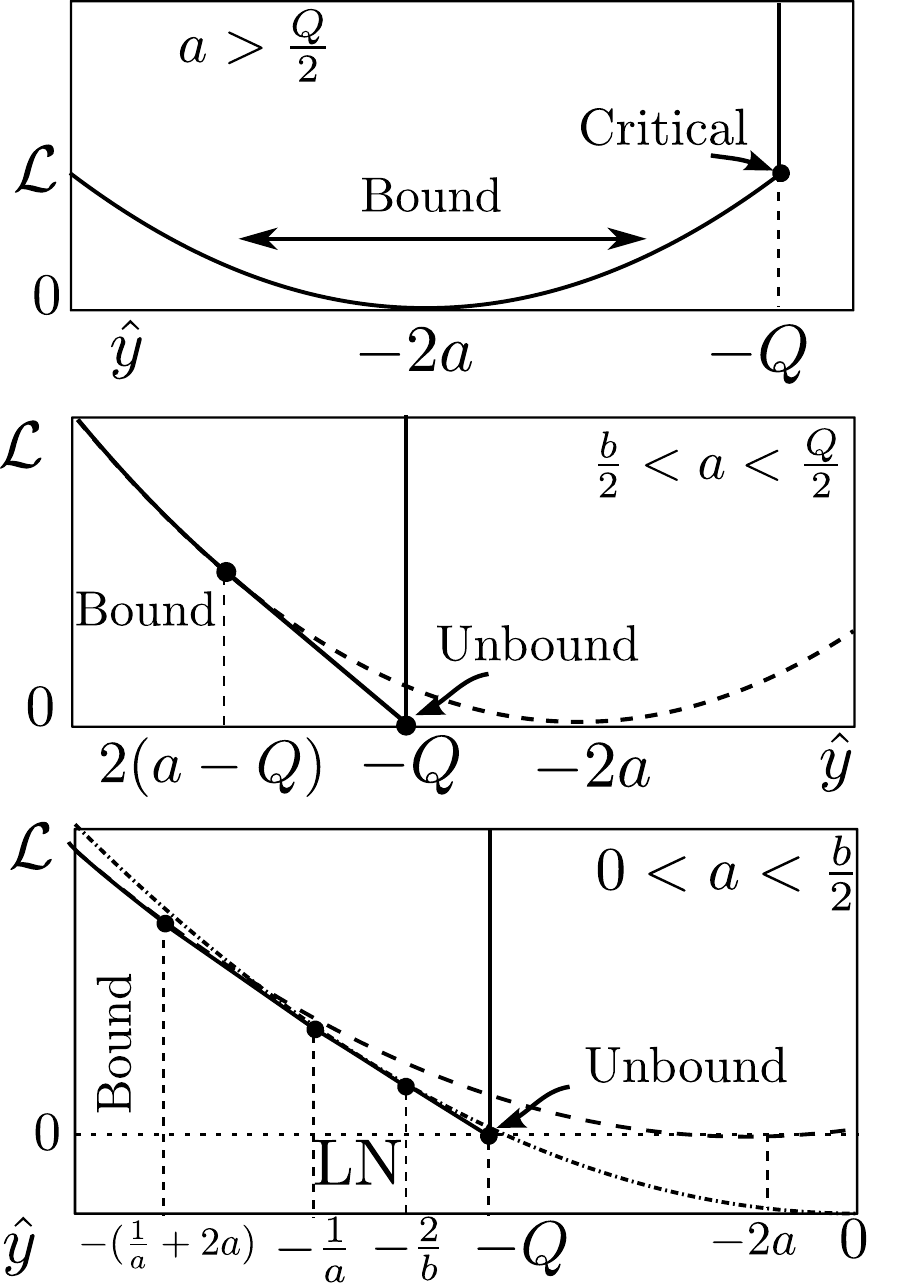} 
	\caption{Illustration of the large deviation function of the free energy $F_a$ of logREMs with one charge. The three panels correspond to eqs. \eqref{eq:ldf1},  \eqref{eq:ldf2} and \eqref{eq:ldf3}, from top to down. The large deviation function $\mathcal{L}(\hat{y} = F_a/ \ln M)$ eq.~\eqref{eq:Lf} is drawn in solid curve. Dots on the curve indicate a non-analyticity.} \label{fig:ldf}
\end{figure}

To better understand the results from the viewpoint of logREMs with one charge, we consider the large deviation function of its free energy $F_a = -\beta^{-1} \ln Z_a$ [$Z_a$ is defined in eq.~\eqref{eq:Za}], defined as:
\begin{equation} \mathcal{L}(\hat{y}) \defeq - \lim_{t\to\infty} \frac1{t} \ln \overline{\delta(F_a / t - \hat{y})} \,,\, t \defeq \ln M \,. \label{eq:LDF} \end{equation}
In other words, the probability distribution of the free energy density is:
$$ P(F_a / t = \hat{y}) = e^{- t\mathcal{L}(\hat{y}) + o(t)}  \,. $$
As a result, the leading exponent of the characteristic function $\tau(s)$ is the Legendre transform of $\mathcal{L}(\hat{y})$:
\begin{equation}\label{eq:legendre0}
\tau(s)  \defeq \lim_{t\to\infty}\left( \frac1{t} \ln  \overline{e^{s F_a}} \right)  = \max_{\hat{y}} (s \hat{y} - \mathcal{L}(\hat{y})) \,.
\end{equation}
On the other hand, $\tau(s)$ can be related to the exponent $\Delta(a,n)$, thanks to the Girsanov transform~\eqref{eq:girnasov} [together with eq.~\eqref{joint_distr}, note also $e^{sF_a} = Z_a^{-s/\beta}$], as follows:
\begin{equation} \tau(s) =- \Delta(a,-s/\beta) - a^2 \,. \label{eq:tauDelta} \end{equation} 
Applying this to the expressions of ${\Delta}(a, n)$ in Table~\ref{tab:res}, we obtain $\tau(s)$ in the four regimes, that we exhibit in its last last column.
Then we calculate $\mathcal{L}(\hat{y})$ as the Legendre transform of $\tau(s)$ (recall that the inverse of Legendre transform is Legendre transform itself). The results, illustrated in Fig.~\ref{fig:ldf}, will be discussed in the Bound and Unbound phases below.

\subsubsection{Bound phase: $a > Q/2$}\label{sec:LDF1B}
\begin{subequations}\label{eq:Lf}
The large deviation function is as follows:
\begin{equation}
\mathcal{L}(\hat{y}= F_a/t) = \begin{dcases}
 \frac14 (\hat{y} + 2a)^2  \,,\, &  \hat{y} < -Q \\
  +\infty  \,,\, &  \hat{y} > -Q 
\end{dcases} \,. \label{eq:ldf1}
\end{equation}
As a consequence, the typical behavior of the free energy is governed by the Bound phase, and is a Gaussian with the following mean and variance:
\begin{equation}  \overline{F_a} = -2a t + o(t), \text{Var}(F_a) =  -2 \ + o(t), t = \ln M \,. \label{eq:FatypBound}  \end{equation}
 Note that the mean value agrees with eq.~\eqref{eq:binding}, while the variance diverges as $M \to \infty$. This extensive variance is a characteristics of the Bound phase: in comparison, $\text{Var}(F_0)$ remains of order unity in logREMs without charge. It is possible to calculate the non-Gaussian corrections to eq.~\eqref{eq:FatypBound} for specific integrable logREMs, see Appendix~\ref{sec:morrisbound} and Ref.~\cite{cao17fbm}.
 
The above Gaussian tail extends to the whole Bound regime $F_a \in (-\infty, Q t + o(t))$. The Critical regime describes rare realizations where the free energy behaves as if the charge did not exist: $F_a = -Q t+ o(t)$.  The probability of $\hat{y} > -Q$ is vanishing even under the large deviation scaling. Indeed, it is rigorously known~\cite{ding2013exponential} that the distribution of the free energy of logREMs without charge has generically a double exponential right-tail beyond its typical value $F_0 = -Qt + o(t)$; since the log-potential is attractive, the free energy of logREMs with one charge is more negative $F_a < F_0$, so we always have a hard wall at $\hat{y} = -Q$. In this sense, $\hat{y} = -Q$ is also a {termination point}, and the C/B transition another kind of termination point transition. We remark that eq.~\eqref{eq:ldf1} was already announced in Ref.~\cite{cao17fbm} [in a slightly different form, see eq. (47) therein], and the termination point transition predicted here played a crucial role in that work, see Appendix~\ref{sec:morrisbound} for further discussion on this matter.
  
\subsubsection{Unbound phase: $a < Q / 2$} 
When $a < Q/2$,  the typical free energy is governed by the Unbound phase, and has the same leading behavior as the free energy of logREMs without charge [see eq.~\eqref{eq:binding}]. However, the negative large deviations of $F_a$ have a highly non-trivial structure. When conditioned to atypically negative free energy, the system can transit from the Unbound regime to the Log-Normal and/or Bound regimes. These are generically first-order transitions, i.e., the derivative $\tau'(s)$ has a jump at the transition. Via Legendre transform, such jumps correspond to intervals where $\mathcal{L}(\hat{y})$ is linear. In what follows we use the short-hand
\begin{equation}
b = \min(\beta , 1) = \begin{cases}
\beta & \beta < 1 \\ 1 & \beta > 1
\end{cases} \label{eq:defb}
\end{equation}
and $Q = b + b^{-1}$ to include both phases, and distinguish two cases:
\begin{itemize}
	\item[-] When $ \min(\beta,1)/2 < a < Q/2$, we have 
	\begin{equation}
	\mathcal{L}(\hat{y}) = \begin{cases}
	(\hat{y} + Q) (2a - Q) &  2(a-Q)  < \hat{y}< -Q \\
	(\hat{y}+ 2a)^2/4 &  \hat{y} < 2(a-Q) 
	\end{cases} \label{eq:ldf2}
	\end{equation}
	The Unbound regime corresponds to a single point $\hat{y} = -Q, \mathcal{L} = 0$, and the Bound regime the parabola of the third line. The first-order transition between the two regimes generates the linear interpolation regime described in the second line of eq.~\eqref{eq:ldf2}. It is obtained as the non-horizontal tangent of the parabola $\mathcal{L} = (\hat{y}+ 2a)^2/4$ that intersects the point $(-Q, 0)$.
	
	\item[-] When $0 <  a < \min(\beta,1)/2 $, there appear two first-order transitions, Unbounded to Log-Normal, and Log-Normal to Bound, as  $\hat{y}$ and $s$ decrease (moving to the right in Fig.~\ref{fig:phase}). So we have a rich alternation of Gaussian and exponential tails:
	\begin{equation}
	\mathcal{L}(\hat{y}) = \begin{dcases}
	-\frac{\hat{y} + Q}b\,, &  -\frac{2}{b}  < \hat{y} < -Q \\ 
	\frac{\hat{y}^2}4 - 1\,,   &  -\frac1a   <  \hat{y} <  -\frac2{b}   \\
	-\frac{\hat{y}}{2a} - 1 - \frac1{4a^2} \,,    &   -\frac1a - 2 a <  \hat{y}  < -\frac1a  \\
	\frac{(\hat{y} + 2a)^2}4 \,, &  \hat{y} <  -\frac1a - 2 a 
	\end{dcases}\label{eq:ldf3}
	\end{equation}
	The two linear parts are found as a tangent line of the parabola $\mathcal{L} = (\hat{y}+ 2a)^2/4$ intersecting $(-Q, 0)$ (the Unbound regime) and a tangent line of both parabolas, respectively. When $a = 0$, the last two cases disappear, and we retrieve the known linear-quadratic tail of the free energy distribution in logREMs without charge~\cite{fyodorov2009statistical}.
\end{itemize}
\end{subequations}

The alert reader may have noticed that the results above rely on the hypothesis that $ \mathcal{L}(\hat{y})$ is convex (otherwise, the Legendre transform of $\tau(s)$ will be the convex hull of $\mathcal{L}(\hat{y})$, but not $\mathcal{L}(\hat{y})$ itself). The convexity assumption is directly responsible for the prediction of the linear parts in $\mathcal{L}(\hat{y})$, which correspond to of exponential tails in the free energy distribution $P(F_a)$. We will check in Appendix~\ref{sec:morris} that these exponential tails can be all found in an integrable logREM, the circular model with one charge, by a completely different method: we calculate the moments $\overline{Z_a^n}$ beyond the leading behavior, by relating to exactly solvable Coulomb gas integrals. By analyzing the pole structure of $\overline{Z_a^n}$, seen as the Laplace transform of the distribution of $F_a$, we identify its exponential tails. They turn out to agree exactly with the linear parts of the large deviation theory prediction,  eqs.~\eqref{eq:Lf} . This is a non-trivial test of the convexity assumption, which we will use again in Section~\ref{sec:LDF2}.


\subsection{Joint large deviation of logREM without charge} \label{sec:LDF2}
Another way to interpret the results in Table~\ref{tab:res} is to consider the joint large deviation function of $-\ln p_{\beta,1}$ and $F_0$ of the logREM \textit{without charge}:
\begin{align}
&P( -\beta^{-1} \ln p_{\beta,1} = t \hat{x}, F_0  = t\hat{y})  \nonumber \\ \defeq
 &\exp\left[t f(\hat{x},\hat{y}) + o(t)\right] \,,\, t \defeq \ln M \label{eq:multifracdef}
\end{align}
which can be also viewed as a generalization of the multi-fractal spectrum of the Gibbs probability weight. We shall consider it for $\hat{y} \leq -Q$ and $\hat{x} \geq 0$ (since $p_{\beta,j}\leq1$ by normalization). Eqs \eqref{eq:multifracdef}, \eqref{eq:girnasov} and \eqref{eq:hmndef} imply then that $f(\hat{x}, \hat{y})$ is related to leading exponent $\Delta (a,n)$ in Table~\ref{tab:res} by Legendre transform:
\begin{equation}\label{eq:legendre1}
\Delta = \min_{\hat{x}, \hat{y}}\left[ a \hat{x}  + (a + n \beta) \hat{y}  - f(\hat{x}, \hat{y}) \right] \,,
\end{equation}
\textit{i.e.}, the variables $\hat{x}, \hat{y}$ are dual to $a, a + n \beta$ (which are the two axes in Fig.~\ref{fig:phase}) respectively. Performing the (inverse) Legendre transform, we obtain the following results: 
\begin{itemize}
\begin{subequations}\label{eq:multifrac}
\item[-] The Unbound regime transforms to
\begin{equation}\label{eq:multifracU}
f(\hat{x} > 0, \hat{y} = -Q) = -\frac14 (\hat{x}- Q)^2 \,,
\end{equation}
which describes realizations where $F_0 \sim -Q t + o(t)$ has its typical value [eq. \eqref{eq:freezing}] and $p_{\beta, 1} \ll 1$. Note that the usual multi-fractal spectrum of the Gibbs probability weight~\cite{chamon1996localization,castillo97dirac,evers2008anderson} is given as $\tilde{f}(\gamma) = f(\hat{x} = \gamma / \beta, -Q) + 1$; the $+1$ difference is due to the fact that $\tilde{f}$ counts all the sites $p_{\beta,j}, j = 1, \dots, M$, whereas eq.~\eqref{eq:multifracdef} describes the distribution of $p_{\beta,1}$ at a single site. We remark also that values of $\hat{x}$ such that $f(\hat{x}, -Q) + 1 < 0$ is absent in a typical large realization, and occurs only in rare samples (this leads to different transitions in the \textit{typical/quenched} ensemble~\cite{evers2008anderson,fyodorov2009pre} of Gibbs measure multi-fractality, whereas our study corresponds to the\textit{ annealed} ensemble). 
\item[-] The Critical regime transforms to a single point
\begin{equation}\label{eq:multifracC}
f(\hat{x} = 0, \hat{y} = -Q) = -\frac{Q^2}{4} \,,
\end{equation}
which describes realizations where $F_0$ is still typical but the Gibbs probability weight at site $1$ is \textit{atomic} $p_{\beta,1} \sim O(1)$. This happens for a few $p_{\beta,j}$'s (amongst $j = 1, \dots, M$) in a typical sample when $\beta > 1$ [because $Q = 2 \Rightarrow f + 1 = 0$, see eq.~\eqref{eq:freezing}], and only in rare samples when $\beta < 1$. Compared to the Bound regime (see below), the Critical regime describes realizations at the onset of a binding to site $1$.
\item[-] The Bound regime transforms to
\begin{equation}\label{eq:multifracB}
f(\hat{x} = 0, \hat{y}< -Q) = -\frac{\hat{y}^2}4 \,,
\end{equation}
which describes realizations with an atomic Gibbs probability weight $p_{\beta,1} \sim O(1)$ and a negative large deviation of the free energy $F_0 \ll -Qt$. Such large deviations are due to an atypically negative potential value $\phi_1 \sim - \hat{y} t$ . From the viewpoint of the logREM without charge, the whole Bound regime describes rare samples. The Girsanov transform eq.~\eqref{eq:girnasov} can be seen as a biased sampling that makes them typical in the logREMs with one charge $a > Q/2$.
\end{subequations}
\end{itemize}
Notice that the above regimes have only covered the boundaries of the $(\hat{x}, \hat{y})$ parameter space. Its interior is occupied by the Log-Normal regime and the interpolation regimes corresponding to first order transitions between them. We refer to appendix~\ref{sec:fxymore} for further details.  

For the sake of comparison with the traveling-wave equation approach, we note down the transformation between dual variables given by Legendre duality eq.~\eqref{eq:legendre1} inside different regime:
\begin{equation}\label{eq:xyoptimal}
\hat{x},\hat{y} = \begin{cases}
Q - 2a, -Q & \text{Unbound} \\ 
0 , -Q & \text{Critical} \\ 
0, -2 (n \beta + a) & \text{Bound} 
\end{cases}
\end{equation}
We will recover these formulas using the traveling-wave equation approach in the next Section, and re-interpret them in term of positions of a diffusing particle.

\section{Traveling-wave equation approach}\label{sec:kpp}
In this Section we use the traveling-wave equation approach to revisit certain aspects of the diagram obtained so far. Since this approach originates from the study of the Branching Brownian motion (BBM) model, we shall review it in Section~\ref{sec:bbm}. Section~\ref{sec:kpp1} derives the basic analytic result, which will be then analyzed by two methods: the real space analysis (section~\ref{sec:real}),  leads to a direct access to the leading exponents, and can be compared to the large deviation results of Section~\ref{sec:LDF2}; the momentum space analysis (Section~\ref{sec:momentum}) recovers all the LFT predictions, and can be compared  in detail to both precedent approaches.

\begin{figure}
	\includegraphics[width=\columnwidth]{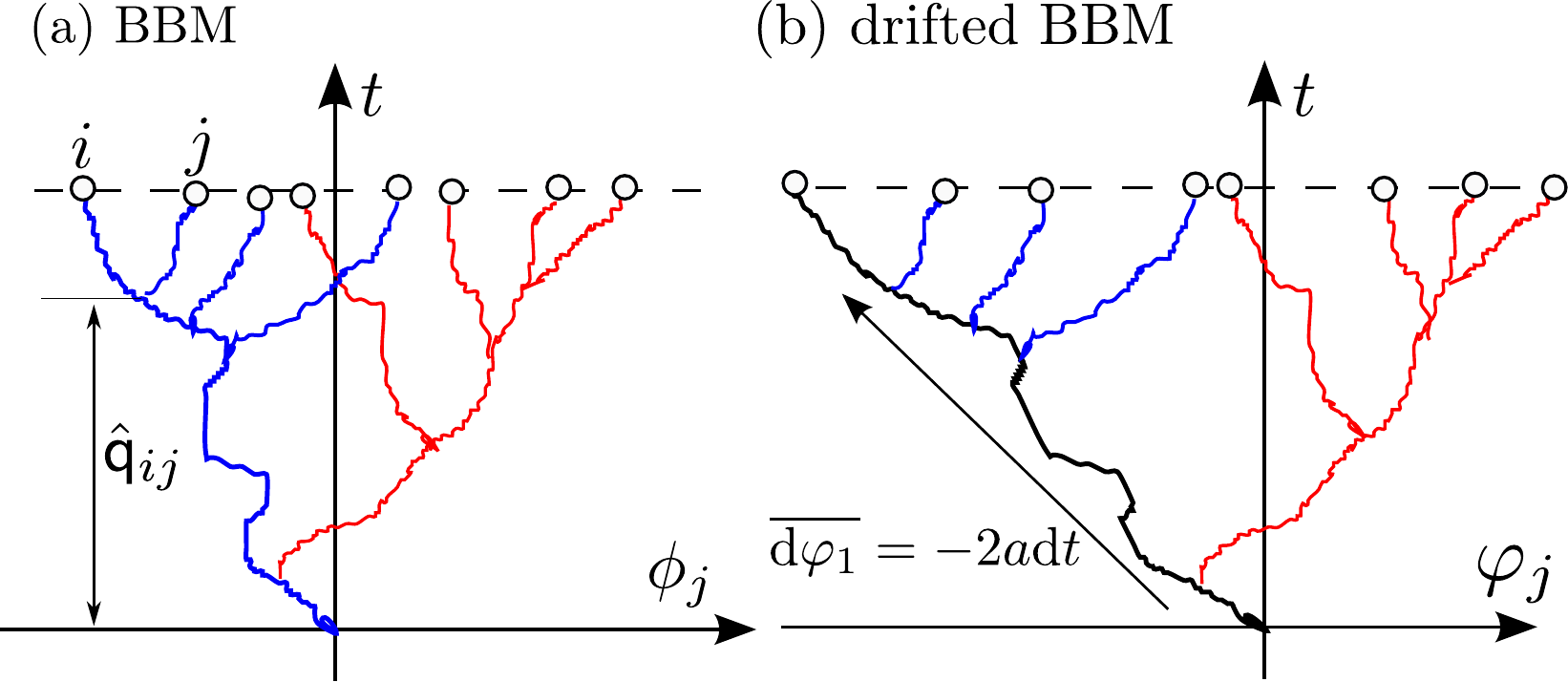}
	\caption{(a) A sample of the BBM model, with the common length illustrated. The two families of particles after the first branching are drawn in different colours. (b) A sample of the drifted BBM. The drifting particle (in black) happens to be the left-most one; note that this is not generally the case.} \label{fig:bbm}
\end{figure}
\subsection{The Branching Brownian motion (BBM) model}\label{sec:bbm}
BBM is a representative of the hierarchical logREMs, and defined by a stochastic process~\cite{makean75bbm,derrida1988polymers}, illustrated in Fig.~\ref{fig:bbm} (a).  The random energy levels $\phi_j = \phi_j(t)$ are the positions (at time $t$) of an ensemble of particles on the real line. The particles diffuse independently with $\overline{\dif \phi_i \dif \phi_j} = 2 \delta_{ij} \dif t $, and each particle splits into two offsprings at the same position with rate $\dif t$ (the splitting events are also independent). Initially, there is only one particle at the origin: $M(t = 0) = 1$, $\phi_1(0) = 0$. As a result, at time $t$, the mean particle number $\overline{M(t)} = e^{t}$ (moreover, $M(t) \sim e^{t}$ typically~\cite{schehr2014exact,ramola2015spatial}), and the partition function defined by eq. \eqref{eq:Z} satisfies eq. \eqref{eq:freezing}. For this reason, the BBM at time $t$ is usually considered equivalent to a logREM of size $M  = e^{t}$: this justifies the notation used since eq. \eqref{eq:defQ}. 
However, there are important {differences} between their definitions. To compare them, let us observe that the random positions of the BBM particles at time $t$ can be generated in two steps. 
\begin{enumerate}
	\item We generate all the branching events up to time $t$, which are independent of the diffusion. This determines the random particle number $M(t)$ and a matrix of \textit{common lengths}, $ \hoverlap_{ij} = \hoverlap_{ij}(t), i,j=1\dots M$, which is most easily defined by an illustration, see Fig.~\ref{fig:bbm}(a). In particular $\hoverlap_{ii} (t) = t$ for any $i$. Note that the common length is closely related to the \textit{overlap}, defined by a rescaling $\overlap_{ij} = \hoverlap_{ij} / t$, see eq.~\eqref{eq:overlap}.
	\item We generate the particle positions $\phi_j = \phi_j(t)$ at time $t$, as correlated Gaussian variables with the following mean and covariance:
	\begin{equation} \overline{\phi_j (t)}^{\hoverlap} = 0\,,\, \overline{\phi_j (t) \phi_i (t)}^{\hoverlap} = 2 \hoverlap_{ij}(t)  \,,\label{eq:commonlength} \end{equation}
   for $i, j = 1, \dots, M(t)$,	where $\overline{[\dots]}^{\hoverlap}$ denotes an average over the diffusion process, while conditioned on a fixed realization of the branching process. 
	By the law of total expectation (a.k.a iterated expectation), generating branching and diffusion in two steps is equivalent to doing both simultaneously. Therefore, the BBM can be seen as a logREM whose covariance matrix is itself random. The total expectation over all BBM randomness will be still denoted by $\overline{[\dots]}$.
\end{enumerate}
It follows from the above observation that BBM belongs to the logREM class if we consider only the typical branching events. When atypical branching events make dominant contributions to an observable, the latter may have different behavior from general logREMs. For example, the large deviation function of the free energy (without charge)  everywhere~\cite{derrida2017large,derrida2017slower}, while there is a hard wall for Euclidean logREMs, see Section~\ref{sec:LDF1}. The difference is precisely due to the contribution of BBM configurations with few or no branching events~\cite{derrida2017large,derrida2017slower}. This issue affects the observables that we study as well and will require a special treatment, as we discuss in detail below.

\subsubsection*{BBM with one charge and drifted BBM}
After reviewing the original BBM model, as a peculiar logREM without charge, we discuss how to define the BBM with one charge (the following discussion is not explicitly required to understand the rest of the paper).  For this, we note that once the covariance matrix eq.~\eqref{eq:commonlength} is generated, the background potential $U_j$ and the composite energy level $\varphi_j = \phi_j + U_j$ can be constructed in the same way by eq. \eqref{eq:defcharge}. Thus, $U_j$ is now random and depends on the branching events. Since the Girsanov relation eq.~\eqref{eq:girnasov} holds for any covariance matrix, it is true for the BBM if $\overline{[\dots]}$ is interpreted as an average conditioned on  any fixed branching event. By the law of total expectation, we conclude that the Girsanov transform eq.~\eqref{eq:girnasov} holds when the average is over all the BBM randomness.

It is interesting to note that the composite energy levels $\varphi_j$ can be also dynamically generated, by a BBM \textit{with one drifting particle} (\textit{drifted BBM}), defined as follows:
\begin{itemize}
	\item[-] The initial particle $\varphi_1$ is a drifting particle, doing a biased Brownian motion: $\dif \varphi_1 =  -2 a \dif t + \dif \phi_1 \,. $
	\item[-] A drifting particle branches into a drifting offspring and a non-drifting one, so that there is always exactly one drifting particle, whose position is denoted $\varphi_1$. 
\end{itemize}
An illustration can be found in Fig.~\ref{fig:bbm}(b). Let us convince ourselves that the particle positions at time $t$ has the same statistical properties as $\varphi_j = \phi_j + U_j$ as defined in eq. \eqref{eq:defcharge}. First, the new definition does not alter the branching events, and thus the statistics of the particle number $M(t)$ and the overlaps $\overlap_{ij}, i,j=1, \dots,M$. Taking the latter as fixed, the above dynamical rule implies that the drift of any particle is given by its common length with particle $1$, so $$U_j = \overline{\varphi_j}^{\hoverlap} = -2 a \hoverlap_{j1} = -a \, \overline{\phi_j \phi_i}^{\hoverlap} $$ by eq. \eqref{eq:commonlength}. This agrees with the definition eq.~\eqref{eq:defcharge}. By the law of total expectation, we conclude that the positions $\varphi_j$ of the drifted BBM are statistically identical to the construction of logREM with 1 charge $\varphi_j = \phi_j + U_j$.

\subsection{From KPP equation to diffusion with absorption}\label{sec:kpp1}
We first recall the fundamental relation between the KPP equation and the BBM~\cite{makean75bbm,derrida1988polymers} (the derivation is recalled in Appendix~\ref{sec:kppderivation}). The exponential generating function
\begin{equation}\label{eq:Gytdef}
G(y,t) \defeq \overline{\exp\left( -e^{\beta y} Z_0 \right)} 
\end{equation}
of the BBM partition function satisfies the following Fisher-KPP equation:
\begin{equation}
G_t = G_{yy} + G(G-1) \,,\, G(y, 0) = \exp(-e^{\beta y}) \,.  \label{eq:kpp}
\end{equation}
It is well-known that when $t \to \infty$, the solution of eq.~\eqref{eq:kpp} tends to a traveling wave, whose position $r(t)$ coincides with the universal asymptotic behavior of the logREM free energy [compare eq.~\eqref{eq:aoft} to eq.~\eqref{eq:freezing}]:
\begin{subequations}\label{eq:kppsol}
\begin{align}
& G(y + r(t),t) \stackrel{t\to\infty}{\longrightarrow} g(y) \,,\,  \label{eq:travel}\\
& \text{where } r(t) = \begin{cases}
-Qt + O(1) & \beta < 1  \\ -Qt + \frac32 \ln t + O(1) & \beta > 1
\end{cases} \,,  \label{eq:aoft} \\
& \text{and $g$ satisfies }g'' - Q g' + g(g-1) = 0 \,,\,   \label{eq:profilebbm}
\end{align} 
\end{subequations}
with limit conditions $g(-\infty) \to 1 \,,\, g(+\infty) \to 0$. In eq.~\eqref{eq:aoft}, $O(1)$ denotes some order-unity quantity. The function $g(y)$ describes the limit profile of the traveling wave solution, and determines the limiting distribution of the free energy $F_0$ for BBM. Indeed, eq. \eqref{eq:Gytdef} implies that $1-G(y)$ is the cumulative distribution function of the convolution $F_0 - \mathsf{Gum}/\beta$, where $\mathsf{Gum}$ is a standard Gumbel random variable independent of $F_0$. In particular, 
\begin{equation} G(y,t), g(y)  \in [0,1] \,. \label{eq:grange} \end{equation} 
and are decreasing function of $y$.

$G(y,t)$ contains only information on the free energy distribution. To access the Gibbs probability weight, we introduce the following observable:
\begin{equation}
H(y,t)  \defeq \overline{\exp\left(-a (\phi_1-y) -e^{\beta y} Z_0 \right)}  \,.
\end{equation}
It is related to the observable of eq.~\eqref{eq:girnasov} by a Laplace transform:
\begin{equation}
\frac\beta{\Gamma(-n)} \int_{\R}  \dif y H(y,t) e^{-(n\beta + a) y} =\,
\overline{Z_0^{n} e^{-a\phi_1}}  \,, \label{eq:laplace}
\end{equation}
where the right hand side is the main observable introduced in eq.~\eqref{eq:girnasov}, evaluated in the BBM model. The above integral converges if and only if $n < 0$, so we shall restrict to this domain in the following. Note that the same restriction applies to the LFT approach in Section~\ref{sec:LFT}.

One can show that the following traveling-wave equation holds for $H(y,t)$ (the derivation, similar to that of eq.~\eqref{eq:kpp}, can be found in Appendix~\ref{sec:kppderivation}):
\begin{equation}
H_t = H_{yy} + (G-1)H \,,\, H(y,0) =  \exp\left( ay - e^{\beta y} \right) \,. \label{eq:kppH}
\end{equation}
This equation is linear in $H$, and can be interpreted in terms of a \textit{single} diffusing particle $y(t)$ with absorption. Its starting position $y(0)$ is distributed with a (non-normalized) probability density $H(y,0)$.  Its diffusive rate is $\overline{(\dif y)^2} = 2 \dif t$ and the absorption rate is $(1 - G(y(t),t)) \dif t$. By eq.~\eqref{eq:laplace} with $(n\beta + a) = 0$, $\overline{p_{\beta,1}^{a/\beta}}$ is proportional to the survival probability of the particle at time $t$. More generally, we have  $ \left< e^{-(n\beta + a) y(t)} \right> \propto \overline{Z_0^{n} e^{-a\phi_1}} $, where $\left< \dots \right>$ denotes the average over the diffusion process times the survival probability.

However, for a reason that we expose in a moment, the asymptotic behavior of eq.~\eqref{eq:laplace} as $t \to \infty$ is \textit{different} from that of Euclidean logREMs with $M = e^t$. For example, let us consider the $(n\beta + a) = 0$ case. Since $G \geq 0$ [eq.~\eqref{eq:grange}], we can bound the term $(G-1)H > -H$, and obtain the following estimate for $\int H := \int_{\R} H(y,t) \dif t $
\begin{equation} \frac{\dif }{\dif t} \int H = -\int  (1-G)H  \geq -  \int H \Rightarrow 
 \int H  \gtrsim  e^{-t} \,. \label{eq:bound} \end{equation}
By eq. \eqref{eq:laplace}, this entails that the exponent $\Delta_{\text{BBM}}(a, n=-a/\beta)\leq 1$ when $(a+n\beta) = 0$, which disagrees with eq.~\eqref{eq:lftdelta}, since $Q^2/4 > 1$ for any $\beta < 1$. The origin of this discrepancy is that, the LFT and RSB methods apply to logREMs with fixed and large size $M$, whereas the above traveling-wave equations describe the BBM model at fixed time $t$, at which the particle number $M(t)$ fluctuates. In particular, there is probability $e^{-t}$ that the initial particle never splits until time $t$, in which case $M(t)=1$ and $p_{\beta, 1} = 1$. Therefore, we can bound from below the annealed average $\overline{p_{\beta,1}^{a/\beta}} \geq e^{-t}$, and deduce that $\Delta_{\text{BBM}}(a, n=-a/\beta)  \leq 1$ for any $a > 0$. 
To recover the exponents which describe logREMs with size $M = e^t \gg 1$, we need to suppress by hand the anomalous events where $M(t) \ll e^{t}$. For this, we propose a simple heuristic approach, which consists in replacing the soft absorbing potential $(1 - G)$ by a hard absorbing wall at the wave-front location $r(t) $ [see eq.~\eqref{eq:aoft}]. The reasoning behind is that, if the maximum value of $(1-G)$, i.e., the maximal absorption rate, is increased to $U > 1$, the estimate eq.~\eqref{eq:bound} will become $ H  \gtrsim  e^{- U t} $. By letting $U \to +\infty$, we will totally suppress these undesired events. Therefore, by analogy with eq.~\eqref{eq:laplace}, we consider the observable 
\begin{equation}
\mathcal{O} \defeq \int_{-\infty}^{r(t)} h(y,t) e^{-(n\beta + a) y} \dif y \,, \label{eq:O}
\end{equation}
where $h$ is defined by the following PDE with moving Dirichlet boundary condition:
\begin{align}
&h_t = h_{yy} \,, y < r(t) \,, h_{y \geq r(t)} = 0 \,,\, \nonumber\\
 & h_{t = 0}(y) = \begin{cases} e^{a y} & y < 0  \\ 0 & y \geq 0 \end{cases} \,. \label{eq:h}
\end{align}
That is, we replace the PDE eq.~\eqref{eq:kppH} for $H$ by its ``hard-wall'' version. We expect that the resulting observable $\mathcal{O}$ reproduces the asymptotic behaviors of $\overline{Z_0^{n} e^{-a\phi_1}}$ for logREMs of size $M = e^t$.

Eq.~\eqref{eq:h} has a even simpler statistical interpretation: it describes a particle that diffuses freely in the half line $y < r(t)$, with Dirichlet boundary condition at $y = r(t)$. At the leading order, $r(t) \approx -Qt$ by eq.~\eqref{eq:aoft}. This brings us to consider the diffusion kernel (from $y(0) = x$ to $y(t)=y$) in presence of a moving wall at $-Q t$, which can be obtained by a mirror image trick. We claim that the result is:
\begin{equation}
D(y,x\vert t) = \frac{1}{\sqrt{4\pi t}} \left[ e^{-\frac{(x-y)^2}{4t}}-  e^{-Qx - \frac{(x+y)^2}{4t}}\right]\,, \label{eq:kernel}
\end{equation}
for any $x < 0$ and $y < -Qt$. Indeed, it can be checked that $D(y,x\vert t)$ satisfies the diffusion equation $D_{t} = D_{yy}$ and the boundary condition $D(-Qt,x\vert t) = 0$. In terms of this kernel, the observable is approximated as [using eq.~\eqref{eq:O} and the initial condition in eq.~\eqref{eq:h}]
\begin{equation}\mathcal{O} \sim \int_{-\infty}^{-Qt}    \dif y \int_{-\infty}^0 \dif x \, D(y,x\vert t) e^{a x - (a + n \beta) y} \,. \label{eq:Oformula}\end{equation}
  Since the approximation consisted in replacing $r(t)$ by $-Qt$, by eq.~\eqref{eq:aoft}, this expression is always exact to the leading order, and to the log-correction order in the $\beta < 1$ phase. Eq.~\eqref{eq:Oformula} will be analyzed in two ways in the following sections.

\subsection{Real space analysis} \label{sec:real}
\begin{figure}
\includegraphics[width=\columnwidth]{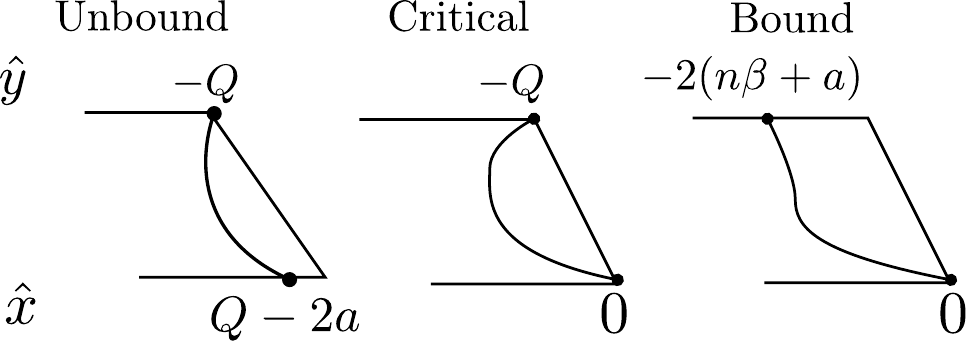}
\caption{Illustration of dominant contributions to the observable $\mathcal{O}$, eq. \eqref{eq:defDeltayy}, in terms of a single diffusive particle with an absorbing wall. The initial and final positions correspond to the log of the Gibbs probability weight at site $1$ and the free energy of logREMs without charges, respectively, see eq.~\eqref{eq:y0ytmeaning} [compare also to eq.~\eqref{eq:xyoptimal}].} \label{fig:bm}
\end{figure}
In this analysis we focus on the leading exponent of the observable $\mathcal{O}$. For this, we write $D(x,y\vert t) = \exp\left[-\frac1{4t} (y - x)^2 + O(\ln t)\right]$ (the mirror image term can be discarded, because it is always sub-dominant), and replace the integrals by an optimization of the endpoint positions $x = -\hat{x} t < 0$, $y = \hat{y} t < r(t) = Qt + o(t)$:
\begin{align} 
&\mathcal{O}  = \exp\left[-t \min_{\substack{ \hat{y} \leq -Q \\ \hat{x} \leq 0}}  \Delta(a,n\vert \hat{x},\hat{y}) \right] \times \text{corrections} \,, \label{eq:defDeltayy} \\
& \text{where }{\Delta}(a,n\vert  \hat{x},\hat{y}) = \frac14 (\hat{y}  + \hat{x})^2 + a \hat{x} + (a + n\beta) \hat{y} \,. \label{eq:Deltayy}
\end{align}
Like $\Delta(a,n\vert m,n_0)$ in eq.~\eqref{eq:hmn}, ${\Delta}(a,n\vert  \hat{x},\hat{y}) $ depends on the variational parameters $ \hat{x},\hat{y}$ to be optimized. One can check that the gradient of ${\Delta}(a,n\vert \hat{x},\hat{y})$ never vanishes in the optimization domain, provided $n < 0$. Hence the minimum is always on the boundary $ \set{\hat{y} = -Q} \cup \set{\hat{x} = 0}$. An explicit calculation shows that the Unbound, Bound and Critical regimes are characterized by the qualitative nature of as summarized in Fig.~\ref{fig:bm}. Plugging the optimal positions into eq.~\eqref{eq:Deltayy} gives the same exponents as in eq.~\eqref{eq:Delta}, or in Table~\ref{tab:res}. 
Moreover, the optimal values of $\hat{x}$ and $\hat{y}$ are identical to the expressions of eq.~\eqref{eq:xyoptimal}, obtained by the two-variable Legendre transform of the RSB results. This justifies the choice of the variable names, and provides clear interpretations of the initial and final positions by a comparison to eq.~\eqref{eq:multifracdef}. Indeed they are respectively the magnitude of the Gibbs probability weight at site $1$, and the free energy of the logREM without charge:
\begin{subequations}\label{eq:y0ytmeaning}
\begin{align} 
&y(0) = -\hat{x} t = \beta^{-1} \ln p_{\beta, 1} \leq 0 \,, \\
& y(t) =  \hat{y}t = F_0 \leq -Qt + o(t) \,.
\end{align}
\end{subequations}

\subsection{Momentum space analysis}\label{sec:momentum}
In order to recover the log-corrections predicted by LFT [eq.~\eqref{eq:eta}] and further connect to the RSB and LFT methods, we analyze eq.~\eqref{eq:h} by a momentum space method. As in the LFT approach, we shall assume $n < 0$ and restrict to the high temperature phase $\beta < 1$. The latter assumption makes the approximation $r(t) \approx -Qt$ leading to eq.~\eqref{eq:Oformula} exact to the log-correction order, so that we can calculate correctly log-corrections (they are known to be different in the $\beta> 1$ phase, and are partially considered in Ref.~\cite{cao17fbm}). 

To obtain the momentum space representation of eq.~\eqref{eq:Oformula}, we apply the following Hubbard-Stratonovich (HS) transform to $D(y,x\vert t)$:
 \begin{equation}\label{eq:HSforD}
 D(y,x\vert t) = \int_{\mathcal{C}} \frac{\dif \alpha}{2 \pi \im} e^{\alpha^2 t}   \left[ e^{(y-x) \alpha} +  e^{ (y+x) \alpha - Q x}  \right] \,,
 \end{equation}
where the integral contour can be any vertical axis in the complex plane.  Plugging eq.~\eqref{eq:HSforD} into eq.~\eqref{eq:Oformula} and integrating over $x$ and $y$ gives:
\begin{subequations}\label{eq:HS1}
\begin{align}
&\mathcal{O} = \int_{\mathcal{C}_1}  \frac{\dif \alpha}{2 \pi \im}  C_1(\alpha)  e^{-t \left[ \Delta_\alpha - Q (a + n \beta) \right] } \,,\, 
 a > \frac{Q}2   \,, \label{eq:HS}  \\
& C_1(\alpha) \defeq \frac{(2 \alpha- Q) }{(\alpha - a) (Q - \alpha - a) (\alpha - a - \beta  n)} \label{eq:Calphay}
\end{align}
where $\Delta_\alpha = \alpha (Q - \alpha)$ [eq.~\eqref{eq:dimensionLFT}] and the integral contour is vertical (oriented towards $+ \im\infty$, same below) and satisfies:
\begin{equation} \mathcal{C}_1: \, \max(Q-a, a + n \beta)< \Re(\alpha) < a \,. \label{eq:domainy} \end{equation} 
\end{subequations}
There is another HS transform formula for $D(y,x\vert t)$, which is the same as eq.~\eqref{eq:HSforD}, except that the term $e^{ (y+x) \alpha - Q x} $ is replaced with  $ e^{ -(y+x+2Qt) \alpha + Q y + Q^2 t}  $. This leads to an alternative momentum-space expression:
\begin{subequations}\label{eq:HS2}
\begin{align}
&\mathcal{O} = \int_{\mathcal{C}_2}  \frac{\dif \alpha}{2 \pi \im}  C_2(\alpha)   e^{-t \left[ \Delta_\alpha - Q (a + n \beta) \right] } \,,\, 
 n \beta + a < \frac{Q}2 \,. \label{eq:HSx} \\
& C_2(\alpha) \defeq \frac{(2 \alpha- Q ) }{(\alpha - a) (\alpha - a - n \beta) (Q-\alpha -a-\beta  n)} \label{eq:Calphax}
\end{align}
where the contour is vertical and satisfies:
\begin{equation} \mathcal{C}_2:\,  a + n \beta < \Re(\alpha) < \min(a,Q - a - n \beta) \,. \label{eq:domainx} \end{equation} 
\end{subequations}
The two sets of formulas cover the entire region $n < 0$ and agree with each other when they can be compared. 

Now, the leading and sub-leading exponents of the observable $\mathcal{O}$ can be calculated by the saddle-point (steepest-descent) approximation. This requires displacing the integral contour to the saddle-point contour:
\begin{equation}  \alpha \in \frac{Q}{2} + \im \R \,. \label{eq:spectrum} \end{equation}
 In the same manner as  in LFT~\cite{cao16liouville}, one must examine whether a pole of $C_1(\alpha)$ or $C_2(\alpha)$ is crossed in the displacement, and take into account the discrete term generated by pole-crossing. The results are as follows:

\begin{itemize}
		\item[-] In the Unbound regime ($a < Q/2$), the pole at $\alpha = a$ in eq.~\eqref{eq:Calphax}  is crossed when displacing the contour to eq. \eqref{eq:spectrum}. So a discrete term dominates instead of the continuous term, so there are no log-corrections (here and below,  ``log-correction'' means a power of $t$, which is proportional to the log of the leading behavior $e^{-\Delta t}$): 
	\begin{subequations} \label{eq:pole}
		\begin{equation} \label{eq:poleU}
		\mathcal{O} \sim 
		\frac{(2 a - Q ) e^{t \left( a^2 + Q n \beta  \right) }}{n \beta (Q-2a-\beta  n)} 
		\end{equation}
		\item[-] Similarly, in the Bound regime ($n\beta + a > Q/2$), the pole $\alpha = a + n \beta$ in eq.~\eqref{eq:Calphay} is crossed, giving:
		\begin{equation}\label{eq:poleB}
		\mathcal{O} \sim	\frac{(2 a + 2 n \beta - Q ) e^{ t  (n\beta + a)^2 }}{n \beta (2 a + n \beta - Q)}
		\end{equation}
	\end{subequations}  
    In the above two equations, the pre-factor vanishes when approaching the regime boundary. This is a recurrent signature of the log-corrections in the neighboring regime (and on the boundary), as we show below (see also Appendix~\ref{sec:morriscritical} and Ref.~\cite{cao17fbm}).
	\item[-] In the Critical regime  ($a > Q/2$, $a + n \beta < Q/2$), for either set of formula (both can be used), the contour eq.~\eqref{eq:spectrum} is inside the validity domain eq.~\eqref{eq:domainy} [and eq.~\eqref{eq:domainx}]. At the saddle point, $\alpha = Q/2$, $C(\alpha)$ and $C'(\alpha)$ have a zero. As a result the saddle point approximation gives:
	\begin{equation}
	\mathcal{O} \sim  \frac{e^{t \left[ Q (a + n \beta) - \frac{Q^2}{4}  \right]} t^{-\frac32}}{ \sqrt{4\pi}\left(a-\frac{Q}{2}\right)^2 \left(a+\beta  n-\frac{Q}{2}\right)^2} \;,  \label{eq:32}
	\end{equation}
	\item[-] On the U/C and U/B boundaries, eq. \eqref{eq:HS} and \eqref{eq:HSx} is a non-zero constant, so we have instead a log-correction with exponent $\frac12$ :
	\begin{equation}
	\mathcal{O} \sim 
	\frac{\sqrt{\pi}e^{t \left[ Q (a + n \beta) - \frac{Q^2}{4}  \right]}  t^{-\frac12}}{ n^2 \beta^2} \,. \label{eq:12}
	\end{equation}
\end{itemize}
The above results agree with the LFT predictions eqs.~\eqref{joint_distr},~\eqref{eq:Delta} and \eqref{eq:eta}, provided the correspondence $M = e^t$. The pre-factors above are not universal for all logREMs, yet we believe that their way of divergence/vanishing approaching the transitions are universal. A test of this claim is provided in Appendix~\ref{sec:morris} for an integrable Euclidean logREM, the circular model with charge. Indeed, we will show that the analytically continued Coulomb gas integrals have very similar pole/zero structure compared to eqs. \eqref{eq:Calphax} and \eqref{eq:Calphay}. 

We now compare the forgoing saddle-point analysis to the RSB and LFT approaches. The connection with the RSB approach is best summarized by the following correspondence between the locus $\alpha$ of dominant contribution (i.e., saddle point or pole) and the optimal value of the RSB variational parameter $n_0$:
\begin{equation}
\alpha = n_0 \beta + a \,. \label{eq:RSBKPP}
\end{equation}
This relation can be obtained by matching the exponent in eq.~\eqref{eq:HS} and that of eq.~\eqref{eq:hmn} (with $m = 1$). In light of eq.~\eqref{eq:RSBKPP}, $\alpha = Q/2$ corresponds to the RSB-stationary point eq.~\eqref{eq:RSBsaddle}. Note that the exponent in eq.~\eqref{eq:HS} is \textit{maximum} at the $\alpha = Q/2$ (compared to other values of $\alpha \in \R$), in agreement with the RSB rule of maximizing $\Delta(n_0, m)$ with respect to $n_0$ if $n < 0$. The validity domains eq.~\eqref{eq:domainy} and \eqref{eq:domainx} imply also $n < n_0 < 0$, which is the domain of $n_0$, imposed as another RSB rule. Therefore, the traveling-wave equation approach provides rationale for the new rules concerning the RSB induced by the charge.

The similarity with the LFT approach is even more remarkable. The momentum-space representations eqs.~\eqref{eq:HS1} and \eqref{eq:HS2} are reminiscent of the conformal bootstrap formula for general four-point correlation functions in LFT [see eq.~\eqref{eq:boot}], in several key aspects:
\begin{enumerate}
	\item A continuous integral over $\alpha \in Q/2 + \im \R$, eq.~\eqref{eq:spectrum} is involved.  This is reminiscent of the continuous term in the conformal bootstrap approach to LFT, which is an integral over the LFT operator spectrum $Q/2 + \im \R$.
	\item The ``structure constants'' eq. \eqref{eq:Calphax}, \eqref{eq:Calphay} vanish at $\alpha = Q/2$; this is a generic feature of the Dorn-Otto-Zamalodchikov-Zamalodchikov (DOZZ)~\cite{dorn1994two,zamolodchikov1996conformal} structure constants of LFT [see Appendix~\ref{sec:discrete}, eq.~\eqref{eq:DOZZzero}].
	\item When the parameters cross a critical value,  a pole of the structure constant eqs. \eqref{eq:Calphax}, \eqref{eq:Calphay} cross the spectrum $Q/2 + \im \R$, producing a discrete term. The same mechanism is behind the genesis of LFT discrete terms~\cite{teschner2001liouville,ribault2014conformal,cao16liouville} (see also Appendix~\ref{sec:discrete}).
\end{enumerate}
More specifically to our problem, the dominant value $\alpha$ determined in the traveling-wave equation approach coincides with the dominant conformal-bootstrap internal charge identified in the LFT approach, see eq.~\eqref{eq:alphaLFT}, with the exception of the Bound regime, where $\alpha = Q - a - n \beta$ in LFT while $\alpha = a + n \beta$ in the traveling-wave approach. Nevertheless, these two charges give rise to the same scaling dimension $\Delta_\alpha = \alpha(Q-\alpha)$ [see eq.~\eqref{eq:dimensionLFT}] and correspond in fact to the same primary field in LFT, by the so-called reflection relation~\cite{ribault2014conformal}.

The similarity between the LFT and the traveling-wave equation approach, as well as the diffusion-absorption interpretation, suggests that the LFT features that we have linked to the logREM Seiberg type transitions are already present in a (0+1)-d approximation of LFT (i.e., considering only the imaginary-time evolution of the zero mode of the Liouville field), also known as Liouville Quantum Mechanics, which has numerous connections: the exponential and extremes of a Brownian motion~\cite{comtet1998exponential,schehr2010extreme}, Anderson transition~\cite{balents97susy}, and more recently holographic models~\cite{bagret16LQM}. It will be interesting to find applications of the theory of Seiberg type transitions to these physical systems.

 
\section{Common length (overlap) distribution}\label{sec:overlap}
We now turn to the common length distribution of the BBM model (and hierarchical logREMs in general), defined by sampling independently two configurations according to a same random realization of Gibbs probability weights:
\begin{equation}
P(\hoverlap) \defeq \overline{\sum_{j,k=1}^M p_{\beta,j} p_{\beta,k} \delta(\hoverlap_{jk} - \hoverlap)} \,, \label{eq:defPq}
\end{equation}
where $\hoverlap$ is a real deterministic variable that is unrelated to the random variables $\hoverlap_{jk}$. $P(\hoverlap)$  is related to the leading finite-size correction of the distribution of the overlap ($\overlap = \hoverlap / t$), whose thermodynamic ($t\to \infty$) limit is known to be $\delta(\overlap)$ for $\beta < 1$ and $\beta^{-1} \delta(\overlap) + (1 - \beta^{-1}) \delta( \overlap - 1)$~\cite{derrida1988polymers,arguin2011genealogy}. Yet, the limit distribution of common length is more involved, and investigated only quite recently~\cite{cao16liouville,derrida16kppfinitesize,derrida2017finite}. The purpose of this Section is deriving the predictions of LFT~\cite{cao16liouville} by a traveling wave calculation. 

\begin{figure}
\includegraphics[width=0.8\columnwidth]{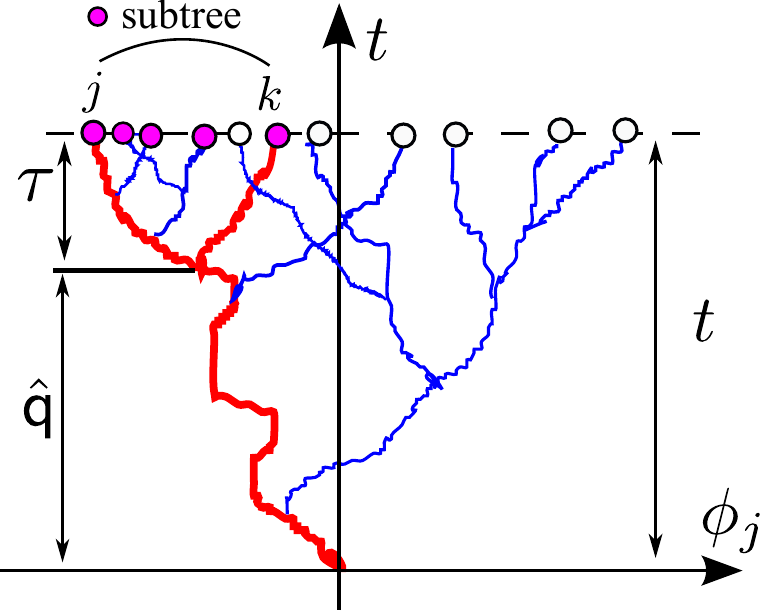}
\caption{Illustration of the observable eq. \eqref{eq:defI} and the time variables $\hoverlap$ and $\tau = t - \hoverlap$. The two marked particles are drawn as red bold curves. The filled particles have common length $\geq \hoverlap$ with $j,k$, and form the ``subtree''. The behavior of its Gibbs measure characterizes the different regimes, see below eq.~\eqref{eq:subtree}. } \label{fig:overlap}
\end{figure}
For this, we adapt the initial step of Section~\ref{sec:kpp} and define
\begin{align}
& I = I(y, \hoverlap, \tau)  \nonumber 
\\  \defeq \, & \overline{\sum_{j,k=1}^{M(t)}  \exp\left(\beta (2y - \phi_j - \phi_k) - e^{\beta y} Z_0 \right)  \delta(\hoverlap_{jk}- \hoverlap)} \,, \nonumber  \\
& \text{where } \phi_j = \phi_j(t) \,,\, \hoverlap_{jk} = \hoverlap_{jk}(t) \,,\, t = \tau + \hoverlap \,.\label{eq:defI}
\end{align}
Here we introduced the variable $\tau = t - \hoverlap$ (see Fig.~\ref{fig:overlap} for illustration), and define $I$ as a function of $y, \hoverlap$ and $\tau$ (instead of $y, \hoverlap$ and $t$), which turns out to be more convenient for writing the traveling-wave equations.

It is not hard to show that $I(y, \hoverlap, \tau)$ is related to $P(\hoverlap)$ by an integral over $y$
\begin{equation}
P(\hoverlap)\vert_{t = \tau + \hoverlap} = \beta \int_{\R} I(y, \hoverlap, \tau) \dif y \,. \label{eq:PandI}
\end{equation} 
In Appendix~\ref{sec:kppderivation}, we show that $I(y, \hoverlap, \tau)$ satisfies the following traveling-wave equation:
\begin{subequations} \label{eq:kppinitI}
\begin{align}
 &I_{\hoverlap} = I_{yy} + (2G(y,\tau+\hoverlap)- 1) I \,,\, \label{eq:kppI}  \\
 &I(y, \hoverlap = 0, \tau) =2 \left(\beta^{-1} G'(y,\tau) \right)^2 \,, \label{eq:initI}
\end{align}
\end{subequations}
where $G(y,t)$ is the solution to the KPP equation, eq.~\eqref{eq:kpp}. Note that $\tau$ is only a parameter of the partial differential equation, and all partial derivatives are with respect to $y$ or $\hoverlap$, $\hoverlap$ being the temporal variable (in the place of $t$).

Eq.~\eqref{eq:kppI} is similar to  eq.~\eqref{eq:kppH}, but has a different quasi-linear term from. Since $(2G-1) \in [-1, 1]$ [recall $g, G\in [0,1]$ in eq.~\eqref{eq:grange}], a direct probabilistic interpretation (as a absorption term) is not possible. Nevertheless, let us consider
\begin{equation}
\tilde{I}(y, \hoverlap, \tau) \defeq  e^{-\hoverlap} I(y, \hoverlap, \tau) \,,\label{eq:Itildedef}
\end{equation} 
which satisfies the following transformed equation
\begin{equation}
\tilde{I}_{\hoverlap} = \tilde{I}_{yy} + 2 (G- 1) \tilde{I} \,, \label{eq:ItildeKPP}
\end{equation}
which enjoys a an interpretation in terms of diffusion with absorption. 

We now make a further simplification by taking the $\tau \to \infty$ limit with $\hoverlap$ fixed. This implies $t = \tau + \hoverlap \to\infty$, so it is a thermodynamic limit. By eq.~\eqref{eq:kppsol},  $G(y,t) \to g(y - r(t))$, $r(t) = -Qt + O(\ln t)$ as $t \to \infty$, where $g(y)$ is the limit profile of $G(y,t)$, see eq.~\eqref{eq:profilebbm}. Since eq.~\eqref{eq:PandI} is not affected by a translation in the $y$-direction of $I$, we may consider 
\begin{equation} I^{\infty} (y, \hoverlap ) \defeq \lim_{\tau\to\infty} \tilde{I}(y - r(\tau), \hoverlap, \tau)  \end{equation} 
Using eq.~\eqref{eq:kppsol} and ~\eqref{eq:ItildeKPP}, we can show that $I^{\infty}$ satisfies the following traveling wave equation
\begin{align}
 &I^{\infty}_{\hoverlap} = I^{\infty}_{yy} + 2(g(y+Q\hoverlap) - 1) I^{\infty} \,, \nonumber \\ 
 & I^{\infty}(y,{\hoverlap = 0}) =  2 (g'(y)/\beta)^2\,, \label{eq:initI1}
\end{align}
On the other hand, using eq.~\eqref{eq:PandI} and eq.~\eqref{eq:Itildedef}, we can relate $I^{\infty}$ to the $\tau \to +\infty$ limit of $P(\hoverlap)$ as follows:
\begin{equation}
P(\hoverlap)\vert_{\tau \to \infty} = \beta e^{\hoverlap} \int_{\R} I^{\infty}(y, \hoverlap, \tau) \, \dif y \,. \label{eq:PandI2}
\end{equation}
A nice consequence of the above equation is an exact {freezing/one-step RSB relation}. Since the traveling-wave velocity $-Q$ and the profile $g$ are independent of $\beta$ in the whole $\beta > 1$ phase [see eq.~\eqref{eq:kppsol} and eq.~\eqref{eq:freezing}], we have, by linearity of eq. \eqref{eq:initI1} and eq. \eqref{eq:PandI2},
\begin{equation}
P(\hoverlap)\vert_{\tau \to \infty, \beta > 1} = \beta^{-1} P(\hoverlap)\vert_{\tau \to \infty, \beta = 1} \,. \label{eq:rsbex}
\end{equation}
In particular, the integral of the left hand side from $\hoverlap = 0$ to $+\infty$ is $1/\beta < 1$. This non-conservation of probability (viewing $P(\hoverlap)$ as a probability distribution) is indeed consistent with the fact that with a finite probability $1 - 1/ \beta$, the overlap $\overlap \sim 1$, which is equivalent to $\hoverlap \sim t \to \infty$: these events escape ``to the infinity'' and are not captured in the $\tau\to\infty$ limit of $P(\hoverlap)$. 
Eq.~\eqref{eq:rsbex} allows us to obtain results in the $\beta > 1$ phase automatically, as long as the $P(\hoverlap \sim O(1))$ regime is concerned (in contrast, the behavior near $P(\hoverlap \sim t)$ is lost in the current setting, the only study of this regime being Refs.~\cite{derrida16kppfinitesize,derrida2017finite}).

Restricting to the $\beta < 1$ phase, we can repeat the probability argument in Section~\ref{sec:kpp}. Indeed, eq.~\eqref{eq:initI1} describes a diffusing particle, in presence of a left-moving soft absorbing wall near $y = -Q \hoverlap$, with absorption rate $2(1 - g(y + Q \hoverlap))$ . In contrast to the situation in Section~\ref{sec:kpp}, the asymptotic behavior will not be affected by replacing the soft wall by a hard one, so our results will apply to the BBM as well as other logREMs. The reason behind this is that the strategy for the particle to stay in the region $y > -Q\hoverlap$ is never optimal. This is because the absorption rate there  $\to 2$ [eq.~\eqref{eq:profilebbm}], so according to eq.~ \eqref{eq:PandI2}, such strategy gives a contribution 
\begin{equation} P(\hoverlap) \sim  e^{-\hoverlap} + \text{other strategies}\,. \label{eq:discard} \end{equation}
Now, it is not hard to see that $P(\hoverlap) \sim e^{-\hoverlap}$ at infinite temperature $\beta = 0$: $e^{-\hoverlap}$ is the common length distribution of two random particles in a BBM, chosen independently and {uniformly} amongst all particles. At any finite temperature, sampling the particles according to a same non-trivial Gibbs measure favors larger common lengths. So $P(\hoverlap)$ decays always slower than $e^{-\hoverlap}$. Since the soft wall is so penalizing, we will consider strategies in which the particle stays always clear of it: $y (\hoverlap) < -Q\hoverlap$.

The analysis of these strategies is very similar to that in Section~\ref{sec:kpp}. More precisely, we can compare $P(\hoverlap)$ to the observable $\mathcal{O}$, defined in eq.~\eqref{eq:O}. The initial distribution of the particle position has an exponential tail $I^{\infty}\vert_{\hoverlap = 0} = 2 (g'(y)/\beta)^2 \sim e^{a y}$ for $y \to -\infty$, where $a = 2 \beta$ [we recall that the left tail $g(y) {\sim} e^{\beta y}, y\to-\infty$ can be seen from the differential equation~\eqref{eq:profilebbm}]. The integral in eq.~\eqref{eq:PandI2} corresponds to $a + n\beta = 0$ in eq.~\eqref{eq:O}. The variable $t$ in eq.~\eqref{eq:O} corresponds to $\hoverlap$ in $P(\hoverlap)$. In summary, we have the correspondence:
\begin{equation} \label{eq:PisO}
P(\hoverlap) \sim e^{\hoverlap} \, \mathcal{O} \left[{a\to 2\beta,n \to -2, t \to \hoverlap} \right]  \,.
\end{equation} 
This allows us to use the results of Section~\ref{sec:momentum}, and find the following regimes: 
\begin{subequations}
\begin{itemize}
\item[-] {Unbound}: When $a  = 2\beta < Q/2 \Rightarrow \beta < 1/\sqrt3$, we apply eq.~\eqref{eq:pole} and obtain:
\begin{equation}
P(\hoverlap) \sim e^{(2\beta^2 - 1)\hoverlap}  \,. \label{eq:Punbound}
\end{equation}
\item[-] {Critical}: $a = 2\beta > Q/2 \Rightarrow \beta > 1/\sqrt3$. Then eq. \eqref{eq:32} applies, and we have:
\begin{equation}
P(\hoverlap) \sim  e^{-(Q^2/4-1) \hoverlap}  \hoverlap^{-3/2} \,.
\end{equation}
At the transition $\beta = 1/\sqrt3$, by eq.~\eqref{eq:12}, the $\overlap^{-3/2}$-correction above is replaced by $\hoverlap^{-1/2}$. 
\item[-] When $\beta \nearrow 1$, the exponent vanishes as $Q^2/4-1 \sim (1-\beta)^2$, leaving only the $\hoverlap^{-3/2}$ term. In the frozen $\beta>1$ phase, by eq. \eqref{eq:rsbex}, this behavior remains:
\begin{equation}
P(\hoverlap) \sim \hoverlap^{-3/2} \,,\, \beta \geq 1 \,,
\end{equation}
in agreement with the prediction of Ref.~\cite{derrida16kppfinitesize}. 
\end{itemize}
\end{subequations}

The above results have been derived by an LFT approach in Ref.~\cite{cao16liouville}. The traveling-wave equation re-derivation here allows to further elucidate their physical meaning. Since we reduced the problem of common length distribution to the same diffusion model as in Section~\ref{sec:kpp}, the interpretation of the dominant initial and terminal positions $y(0)$ and $y(\hoverlap)$ are similar to eq.~\eqref{eq:xyoptimal}. More precisely, $y(\hoverlap) = -Qt + o(t)$ is still the free energy of dominating configurations, whereas  \begin{equation} e^{-\beta y(0)} = \sum_{j \in \text{subtree}} p_{\beta,j} \label{eq:subtree}\end{equation}
is the Gibbs measure of the \textit{subtree}, i.e., the sum of Gibbs probability weight of all sites having common length $\geq \hoverlap$ with the marked pair in eq.~\eqref{eq:defPq}, see Fig.~\ref{fig:overlap}. This interpretation generalizes naturally eq.~\eqref{eq:xyoptimal}, which identifies $e^{-\beta y(0)}$ to the Gibbs probability weight of a single site. 

 In light of the above observation, we remark that in the Unbound regime ($\beta < 1/\sqrt{3}$), the marked pair is effectively attracted by a subtree with Gibbs measure larger than the most typical value $e^{-\beta Q \hoverlap} $, but much smaller than $1$ [see eq.~\eqref{eq:xyoptimal} with $a = 2\beta$]: $$e^{-\beta Q \hoverlap} \ll e^{-\beta y(0)} = e^{(-\beta Q + 4 \beta^2) \hoverlap} \ll 1 \,.$$
In the Critical regime, $\beta \in (1/\sqrt{3},1)$, the Gibbs measure of the subtree becomes of order unity and cannot increase further by normalization: this termination effect is responsible for the Unbound-Critical transition. Since we are still in the $\beta < 1$ phase, such subtrees appear only in rare samples, but their contribution dominates the observable $P(\hoverlap)$. Such subtrees become typical at the freezing transition $\beta=1$, beyond which point $\hoverlap$ becomes ``scale-free'': the leading exponent in $P(\hoverlap)$ vanishes, leaving only the power-law $\hoverlap^{-\frac32}$ in the whole frozen phase [by eq.~\eqref{eq:rsbex}]. We believe that this is a signature of the criticality (or marginal stability~\cite{fyosom2007}) of the frozen phase. 

\section{Conclusion}\label{sec:discussion}
We have studied the universal scaling behavior of moments of the partition function of logREMs in presence of a deterministic logarithmic potential. Although such a model was introduced since the pioneering works on logREMs~\cite{carpentier2001glass}, the scaling behavior of the partition function moments turns out to be quite rich and described by a 2d diagram, in which appear both Seiberg type transitions: binding and termination point (a.k.a pre-freezing). This unified framework allowed us to better understand and generalize main results in previous related works. As a nice corollary, we systematically studied the rich large deviation structure of logREMs with and without charge. Last but not least, our investigation deepened the connection between LFT and universal properties of the logREM class in an unexpected way: the binding transition provides a statistical-physical application of the non-local property of the operator product expansion (OPE) in LFT.   %

Let us close by discussing a couple of interesting perspectives for future work.. Although the LFT predictions have been abundantly corroborated, an important conceptual issue remains: it is highly counter-intuitive that the non-locality of LFT OPE is related to the Seiberg type transitions, which are local in the sense that they are induced by one log-singularity/charge insertion. We believe this ``paradox'' has to do with the fact that our main observable contains an inherently non-local object: the partition function $Z_0$. In a related note, we recall that in the LFT calculation of Section~\ref{sec:mappingLFT}, it is important to consider LFT on the sphere, in order to product the correct power of $Z_0$. On a surface of higher genus, the LFT correlation function corresponding to the observable in eq.~\eqref{joint_distr} would be different. Yet, we expect that the topology of the surface on which we define the logREM should not affect its scaling behaviors. How can we understand such a universality from the  point of view of LFT on general geometries? 

A deeper understanding of the above issue may also be helpful for curing another weakness of the LFT approach developed so far: its inability to study transitions ``in the bulk'' , such as the freezing transition.  Among the LFT-accessible non-trivial regimes, the closest to the $\beta>1$ frozen phase is the Critical regime; for example, the universal log-corrections are remarkably similar. But the Critical regime describes only rare samples or large deviations, so can be only attained by a biased sampling, e.g., from the Unbound regime by favoring realizations in which a privileged Gibbs probability weight $p_{\beta,1}$ becomes of order unity. This procedure breaks the replica symmetry explicitly. In contrast, the freezing transition breaks the replica symmetry spontaneously, without being induced by a charge: in the frozen phase, Gibbs probability weights of order unity (atoms) emerge in {multiple remote} positions of a typical large sample. While it is still not clear how to describe such a proliferation of atoms in a solvable field theory framework, we suspect that some non-local field theory property could be involved.

\begin{acknowledgements}
We thank V. Belavin, Y. V. Fyodorov, D. Ostrovsky, and S. Ribault, for enlightening discussions and collaborations on related projects, and D. Ostrovsky for valuable communications.  X.C. acknowledges support from a Simons Investigatorship, Capital Fund Management Paris, and Laboratoire de Physique Statistique et Modèles Statistiques. PLD acknowledges support from ANR grant ANR-17-CE30-0027-01 RaMaTraF. The research of A.R. is supported by  the ANR grant ANR-16-CE30-0023-01 THERMOLOC and in part by the National Science Foundation under Grant No. NSF PHY 17-48958.	
\end{acknowledgements}

\appendix

\section{Girsanov transform}\label{sec:girsanovapp}
We recall briefly the Girsanov transform for Gaussian random variables and apply it to derive eq.~\eqref{eq:girnasov} and eq.~\eqref{eq:girsanov2}. 

The Girsanov transform is a classic result in probability, which can be stated in general as follows. Let $\mathbf{\phi} = (\phi_1, \dots, \phi_M)$ be a Gaussian vector characterized by zero mean and an invertible covariance matrix $C_{ij} = \overline{\phi_i \phi_j}$. Let ${u} = (u_1, \dots, u_M)$ be a deterministic vector and let $U = C u$. let $\mathcal{F} = \mathcal{F}[{\phi}]$ be an arbitrary function of the vector $\phi$. Then, for any function $\mathcal{F} = \mathcal{F}[{\phi}] $,
\begin{align}
&\overline{e^{- {u}^t {\phi}} \mathcal{F}[{\phi}] } = \int \dif^M \phi \frac{1}{N} e^{- \frac12 \phi^t C^{-1} \phi  - {u}^t {\phi} } \mathcal{F}[{\phi}] \nonumber \\
=& \int  \dif^M \phi \frac{1}{N} e^{- \frac12 (\phi + U) C^{-1}  (\phi + U) + \frac{1}{2} u^t C u } \mathcal{F}[\phi + U]  \nonumber \\ 
=& e^{\frac{1}{2} u^t C u} \overline{ \mathcal{F}[\phi + U] } \,. \label{eq:girsanovgen}
\end{align}
Here, $N = \sqrt{2 \pi \det(C)}$ is a normalization constant.

Now, to derive eq.~\eqref{eq:girnasov}, we let $\phi$ be the logREM potential [eq.~\eqref{eq:logdecay}], and $u = (a, 0, \dots, 0)$, so that $U = C u$ is the log potential eq.~\eqref{eq:defcharge}.  We set also $\mathcal{F}[\phi] = Z_0^n$, so that $\mathcal{F}[\phi + U] = Z_a^n$. Then eq.~\eqref{eq:girsanovgen} implies
\begin{equation}
\overline{ e^{-a \phi_1}   Z_0^{n}} = e^{\frac12 a^2 \overline{\phi_1^2}} \overline{Z_a^{n}}  = M^{a^2} \, \overline{Z_a^{n}}  
\end{equation}
by eq.~\eqref{eq:logdecay}, as desired.

 To derive eq.~\eqref{eq:girsanov2}, we let $\mathcal{F}[\phi] = e^{-a_1 \phi(0) - a_4 \phi(z)} Z_0^n$, $\phi$ be the potential of the logREM on the sphere, and $u = (0,\dots, 0, a_2, 0, \dots, 0, a_3, 0, \dots)$, where $a_2, a_3$ appear at the indexes corresponding to $z=1,z=\infty$, respectively. Then eq.~\eqref{eq:girsanovgen} implies 
 \begin{align}
& \overline{ e^{-a_1 \phi(0) - a_4 \phi(z)} Z_0^n  e^{-a_2 \phi(1) -a_3 \phi(\infty)} } \nonumber \\ = 
 & e^{\frac{1}{2} \left(a_2^2 \overline{\phi(1)^2} + a_3^2 \overline{\phi(\infty)^2} \right)} e^{a_2 a_3 \overline{\phi(1) \phi(\infty)} }  \nonumber \\
 & \times \overline{ e^{-a_1 \tilde\phi(0) - a_4 \tilde\phi(z)} \tilde{Z}_0^n }  \nonumber \\
= &   M^{a_2^2 + a_3^2}  \overline{ e^{-a_1 \tilde\phi(0) - a_4 \tilde\phi(z)} \tilde{Z}_0^n }  \times e^{a_2 a_3 \overline{\phi(1) \phi(\infty)} }  \nonumber \\
\sim & M^{a_2^2 + a_3^2}  \overline{ e^{-a_1 \tilde\phi(0) - a_4 \tilde\phi(z)} \tilde{Z}_0^n }   \nonumber \\
\sim & M^{a_2^2 + a_3^2}  \overline{ e^{-a_1 \phi(0) - a_4 \phi(z)} \tilde{Z}_0^n }   \label{eq:girsanov2derivation}
 \end{align}
where $\tilde{\phi} = \phi + U$ is the shifted potential with a log-potential with two charges at $1$ and $\infty$:
\begin{equation} U(z) = a_2\, \overline{\phi(z) \phi(1)} +a_3\, \overline{\phi(z) \phi(\infty)} \,, \label{eq:U1infty} \end{equation}
and $\tilde{Z}_0 = {Z}_0\vert_{\phi\to \tilde{\phi}}$. 
In the second last line of eq.~\eqref{eq:girsanov2derivation}, we omitted the factor $e^{a_2 a_3 \overline{\phi(1) \phi(\infty)} }$, which remains of order unity as $M\to\infty$ and does not affect the asymptotic behavior; in the last line, we replaced $e^{-a_j \tilde\phi(z_{j})}$ by $e^{-a_j \phi(z_{j})}$ for $j=1,4$, omitting order unity factors $e^{-a_j U(z_j)}$ . Combining eq.~\eqref{eq:girsanov2derivation} with eq.~\eqref{eq:K2} gives eq.~\eqref{eq:girsanov2}. 

We now discuss the assumption made below eq.~\eqref{eq:girsanov2}, i.e.,  
\begin{equation} \overline{ e^{-a_1 \phi(0) - a_4 \phi(z)} \tilde{Z}_0^n }  \sim \overline{ e^{-a_1 \phi(0) - a_4 \phi(z)} {Z}_0^n } \,. \label{eq:assumption} \end{equation} 
For this let us remark that in the Unbound regime, the results eqs.~\eqref{eq:Delta} and \eqref{eq:eta} translate to the following by the Girsanov transform eq.~\eqref{eq:girnasov}:
\begin{equation}  \overline{Z_a^n} \sim M^{Q n \beta} \sim  \overline{Z_0^n} \,,\, a < Q/2, n < 0 \,. \end{equation}
In other words, adding a background log potential with a charge that satisfies the Seiberg bound $a < Q/2$ does not alter the scaling behavior of the partition function of the logREM with charge. Apply this statement to each of the two charges $a_2, a_3 < Q/2$ of the log potential eq.~\eqref{eq:U1infty}, we obtain $\overline{ \tilde{Z}_0^n} \sim  \overline{{Z}^n_0}$. This is weaker than eq.~\eqref{eq:assumption} that we want. However, on intuitive grounds, we expect that the correlation between $\tilde{Z}_0$ (or ${Z}_0$)  and $ e^{-a_1 \phi(0) - a_4 \phi(z)}$ is not altered by the potential eq.~\eqref{eq:U1infty}, which is smooth around $0$.

\section{Conformal bootstrap and discrete terms in LFT}\label{sec:discrete}
In this Appendix, we revisit the analysis of Ref.~\cite{cao16liouville}, Supplementary Material (SM) C3 and C4, and generalize them to take into account the remote discrete terms. 

We recall the main result eq.~(C.14) therein, which is a general conformal bootstrap formula for a general LFT four-point function, with discrete terms included:
\begin{widetext}
	\begin{eqnarray}
	\mathcal{K} =   
	\left< \vertex_{a_1}(0)\vertex_{a_4}(z)\vertex_{a_2}(1)\vertex_{a_3}(\infty)\right>_\beta =  \int_{\frac{Q}{2} + \im \mathbb{R}} C^{\text{DOZZ}}(a_1,a_4,\alpha)C^{\text{DOZZ}}(Q-\alpha,a_2,a_3)|\mathcal{F}_{\Delta_{\alpha}}(\{a_i\},z)|^2\dif \alpha  \nonumber \\
	-   2 \sum_{p \in P_{14,-}} 2\pi \im \;\text{Res}_{\alpha \to p} \left[C^{\text{DOZZ}}( \alpha,a_1,a_4) C^{\text{DOZZ}}(Q - \alpha,a_2,a_3)\right]  \abs{\mathcal{F}_{\Delta_{\alpha}}(\{a_i\},z)}^2  \nonumber \\ 
	-   2 \sum_{p \in P_{23,-}} 2\pi \im \;\text{Res}_{\alpha \to p} \left[C^{\text{DOZZ}}( \alpha,a_1,a_4) C^{\text{DOZZ}}(Q - \alpha,a_2,a_3)\right]  \abs{\mathcal{F}_{\Delta_{\alpha}}(\{a_i\},z)}^2  \,. \label{eq:boot}
	\end{eqnarray}
 The field theory objects involved in this equation are explained as follows. $\mathcal{F}_{\Delta_{a}}(\{a_i\},z)$ is the four-point conformal block. It has the following asymptotic behavior as $z\to 0$:
\begin{equation} \mathcal{F}_{\Delta_\alpha}(\{\Delta_{a_i}\},z) \stackrel{z\to0} \sim z^{\Delta_\alpha-\Delta_{a_1}-\Delta_{a_4}} \,, \label{eq:block}  \end{equation}
where we recall $\Delta_a = a (Q-a)$ is the scaling dimension of $\vertex_a$, see eq.~\eqref{eq:dimensionLFT}. 

$C^{\text{DOZZ}}$ is the Dorn-Otto-Zamolodchikov-Zamolodchikov (DOZZ) structure constant of LFT~\cite{dorn1994two,zamolodchikov1996conformal}. Referring to SM eq.~(C.4) for a detailed exposition, we recall its following properties. For generic $a_1, a_4$, as a function of $\alpha$, both $C^{\text{DOZZ}}(a_1,a_4,\alpha)$ and $C^{\text{DOZZ}}(Q-\alpha, a_2,a_3)$ have a simple zero at $\alpha=Q/2$~\cite{seiberg1990notes,ribault2015liouville}, so that:
\begin{align}
&	C^{\text{DOZZ}}(a_1,a_4,\alpha )  C^{\text{DOZZ}}(Q-\alpha, a_2,a_3) \stackrel{\alpha\to Q/2}{\sim} (\alpha - Q/2)^2  \label{eq:DOZZzero}
\end{align}
$C^{\text{DOZZ}}(a_1,a_4,\alpha)$ and $C^{\text{DOZZ}}(Q-\alpha, a_2,a_3)$ also have simple poles in the pole set $P_{14, +} \cup P_{14, -}$ and $P_{23, +} \cup P_{23, -}$, respectively, where 
\begin{equation} 
P_{ij,-} \defeq  \left\{ x = a_i+ a_j + n \beta + m/\beta : n,m =0,1,2,\dots, x \in  (0, Q / 2) \right\} \,,\, P_{ij,+} 
\defeq \set{x: x = Q - p, p \in P_{ij,-}} \,.  \label{eq:poles}
\end{equation}

Finally, we introduce some less standard terminology. In eq.~\eqref{eq:boot}, we call the summed/integrated variable $\alpha$ the ``{internal charge}'', as in Section~\ref{sec:nonlocal}.  The integral over $ \frac{Q}{2} + \im \mathbb{R}$ (which is the LFT operator spectrum) in the first line of eq.~\eqref{eq:boot} is called the \textit{continuous term}. The residues with respect to the poles in $P_{14,-}$ and $P_{23,-}$ are called the \textit{local} and \textit{remote} \textit{discrete} terms, respectively. {Note that the distinction between local and remote is arbitrary for a general four-point function (since the fields $1,4$ and $2,3$ play exchangeable roles), yet makes sense in the  situation where we analyze the asymptotic behavior when only the fields $1,4$ approach each other.}

With the above information, we can analyze the asymptotic behaviors of the different terms in eq.~\eqref{eq:boot} as $\abs{z}^2 \sim 1/M \to 0$:
\begin{itemize}
	\item[-] The \textit{local discrete term} corresponding to each pole $\alpha \in P_{14,-}$ behaves as 
 \begin{equation}\label{eq:Dis}
  \text{Dis.}(\alpha) \sim M^{\delta} \,,\, \delta = \Delta_{a_1} + \Delta_{a_4} - \Delta_\alpha \,.
 \end{equation}  
   Discrete terms never have log-corrections.  Local discrete terms exist if $P_{14}$ is not empty, which is equivalent to $a_1 + a_4 < Q/2.$ When this is the case, the dominant contribution corresponds to the smallest $\delta$, which is attained by $\alpha = a_1 + a_4$.
	\item[-] The \textit{remote discrete terms} behave also as eq.~\eqref{eq:Dis}, but with $\alpha \in P_{23,-}$. Such discrete terms exist if $a_2 + a_3 < Q/2.$ When this is the case, the dominant contribution is that with $\alpha = a_2 + a_3$.
	\item[-] The \textit{continuous term}'s asymptotic behavior is evaluated by a saddle point approximation of eq.~\eqref{eq:boot}  around $\alpha = Q/2$ [see also SM C.4, case (b) and (c)]:
	\begin{equation}\label{eq:Cont}
	 \text{Cont.} \sim M^{ \delta} \left(\ln M  \right)^{-\eta} \,,\, \delta =\Delta_{a_1} + \Delta_{a_4} -  \Delta_{Q/2} \,, \end{equation}
 where the log-correction has an exponent $\eta = 3/2$ generically, due to the DOZZ zeros at the saddle point $Q/2$, see eq.~\eqref{eq:DOZZzero}. An exception happens when $a_1 + a_4 = Q/2$ or $a_2 + a_3 = Q/2$. In this case $\eta = 1/2$, because the DOZZ zeros become canceled by the DOZZ poles in eq.~\eqref{eq:poles} approaching $Q/2$.
\end{itemize}
The three points above correspond to those below eq.~\eqref{eq:alphaLFT}, respectively. The asymptotic behavior of the four-point function is obtained as that of the dominant contribution among the above candidates. This corresponds to the smallest scaling dimension $\Delta_\alpha = \alpha(Q-\alpha)$, and thus the smallest $\alpha$ (since $\alpha \leq Q/2$ for all contributions). Summarizing the above discussion then leads to eqs.~\eqref{eq:Kasym}, \eqref{eq:deltaLFT} and \eqref{eq:alphaLFT} in the main text.

Note that the remote discrete terms were not considered in Ref.~\cite{cao16liouville}, SM C.4. Nevertheless this does not invalidate the logREM predictions in that work, because as we see from the present work, the remote discrete terms are only present in the Bound regime, which was not studied in Ref.~\cite{cao16liouville}. 

\section{Circular model with one charge}\label{sec:morris}
We consider the main observable eq.~\eqref{eq:girnasov} for an integrable logREM, the circular model~\cite{fyodorov2008statistical}. It is a 1d Euclidean logREM whose potential is defined by restricting the 2d GFF on the unit circle (note that the normalization of the potential below is adjusted to ensure compatibility with eq.~\eqref{eq:logdecay} with $d = 1$):
\begin{align}  
&\phi_j = \phi(z_j) \,,\, z_j \defeq -e^{2\pi \im j / M} \,,\,  
 \overline{\phi(z) \phi(w)} = -2 \ln \abs{z - w} \,,\, z \neq w \,. \label{eq:cirlcularcov}
  \end{align}
In particular, the marked site $j = 1$ corresponds to $z = -1$ (as $M \to \infty$), so the circular model with one charge has as background potential $U(z) = 2a \ln \abs{z + 1}$.
  
The joint moments in eq.~\eqref{eq:girnasov} of the circular model can be evaluated in beyond its leading behavior in the thermodynamic limit, by relating to an exactly solvable Coulomb gas integral. This method is commonly used to study integrable logREMs, and can be understood as an extension of the RSB approach in Section~\ref{sec:rsb} beyond the leading behaviors. It relies on the different RSB solutions in each regime or phase. In the case of the RSB induced by the freezing transition, this method was developed in Refs.~\cite{fyodorov2010freezing,cao16order}, see also Ref.~\cite{cao17thesis}, Section~2.3.3. Adapting the methods therein to the RSB induced by Seiberg type transitions turns out to be quite delicate in some cases. So we will treat the different regimes separately.

 In particular we will recover the exponential tail predicted by the large deviation theory (see Section~\ref{sec:LDF1}) in Sections~\ref{sec:morrisU} and \ref{sec:morrisLN}. Remarks related to Ref.~\cite{cao17fbm} are made in Sections~\ref{sec:morriscritical} and ~\ref{sec:morrisbound}.

	\subsection{Morris integral}
	As a preparation, we recall the Morris integral [our notation in eq.~\eqref{eq:morrisgamma} and eq.~\eqref{eq:morrisbarnes1} below is closest to that of Ref.~\cite{cao17fbm}, eq. (14-15)]:
	\begin{align}
	\mathcal{M}(\tilde{n}, \alpha, b) &\defeq \int_{0}^{2\pi} \prod_{\mu=1}^{\tilde{n}} \left[ \frac{\dif\theta_\mu}{2\pi}  \abs{1 +  e^{\im \theta_\mu}}^{-2\alpha b} \right]
	\prod_{\mu < \nu} \abs{ e^{\im \theta_\mu} -  e^{\im \theta_\mu}}^{-2b^2}  = \prod_{j=0}^{\tilde{n}-1} \frac{\Gamma(1 - 2\alpha b-jb^2)\Gamma(1-(j+1)b^2)}{\Gamma(1 - \alpha b - jb^2)^2 \Gamma(1 - b^2)}  \,. \label{eq:morrisgamma} 
	\end{align}
	It depends on three parameters: the number of moving charges $\tilde{n}$, the total $\alpha$ attached to $z = -1$, and the moving charge value $b$. Eq.~\eqref{eq:morrisgamma} holds if and only if the integral converges, yet the Gamma product in the RHS can be analytically continued beyond its region of convergence and to $\tilde{n}$ complex~\cite{fyodorov2015moments,ostrovsky2016gff}: 
	\begin{align}
	\mathcal{M}(\tilde{n}, \alpha, b) = \frac{ \widetilde{\mathcal{M}}(\tilde{n}, \alpha, b)}{ \Gamma(1-b^2)^{\tilde{n}}} \,,\  	\widetilde{\mathcal{M}}(\tilde{n}, \alpha, b) = \Gamma(1 - \tilde{n}b^2) \frac{\Barnes_b(Q-2\alpha)\Barnes_b(Q - \alpha-\tilde{n}b)^2}{\Barnes_b(Q-2\alpha - \tilde{n}b)\Barnes_b(Q - \alpha)^2} \frac{\Barnes_b(Q)}{\Barnes_b(Q - \tilde{n}b)} \,,\, Q = b+b^{-1} \,,
	\label{eq:morrisbarnes1}
	\end{align}
where $\Barnes_b(x)$ is the generalized Barnes function. We refer to Ref.~\cite{cao17fbm}, eq. 16-18 and references therein for its basic properties. We will not need them, except the following fact: $\Barnes_b(x)$ is an entire function of $x$ that has simple zeros in the following positions:
\begin{equation} \Barnes_b(x) = 0 \,,\, x = -u b - \frac{v}{b}  \,,\, u,v =0,1,2,\dots  \,. \label{eq:barneszeros}\end{equation} 
\end{widetext}

\subsection{Unbound regime}\label{sec:morrisU} 
In the Unbound regime of the $\beta < 1$ phase, the RSB solution (Table~\ref{tab:res}) is $n_0 = 0$ and $m=1$, i.e., the replicas do not form groups and are not attached to the charge. Then, one can apply the covariance eq.~\eqref{eq:cirlcularcov} to the replica sum expression of $\overline{ Z_0^n e^{-a\phi_1} }$, eq.~\eqref{eq:wick}, and replace the sums over positions by integrals on the circle $ \sum_{j_\mu} \leadsto M  \int_0^{2\pi}  \frac{\dif\theta_\mu}{2 \pi} \,. $
The result is proportional to a Coulomb gas integral which coincides with the Morris integral above, with the following parameters~\cite{fyodorov2009pre,fyodorov2015moments,ostrovsky2016gff}:
\begin{align}\label{eq:discont}
 \overline{Z_0^n e^{-a\phi_1} } M^{\Delta(a,n)}  \sim {\mathcal{M}}(\tilde{n} = n,\alpha =  a, b = \beta)  \,,
\end{align}
 where $\Delta(a,n) = -Qn\beta-a^2$ is the leading exponent of the Unbound regime in (see Table~\ref{tab:res}) and the continued Morris integral provides the order-unity, model-dependent correction.  The approximative equality in eq.~\eqref{eq:discont} is expected to be exact in the $M\to\infty$ limit. We remark that a special case of eq.~\eqref{eq:discont} with $a = -n\beta$, see Ref~\cite{fyodorov2015moments}, and Ref.~\cite{cao17fbm}, eq. (13); a further specialization with $n = -2$ appeared in Ref.~\cite{fyodorov2009pre}.

To extract information on the distribution of the free energy of the logREM with one charge $F_a = -\beta^{-1} \ln Z_a$, we apply the Girsanov transform eq.~\eqref{eq:girnasov} to eq.~\eqref{eq:discont}, and obtain the moment generating function of the shifted free energy~\cite{ostrovsky2016gff}:
\begin{equation}
\overline{e^{sf_a}}  \sim {\mathcal{M}}(-s/\beta, a, b = \beta) \,,\, f_a \defeq F_a + Q t \label{eq:expFaMorris}
\end{equation}
 Since $\overline{e^{sf_a}}$ is the Laplace transform of the distribution $P(f_a)$, the poles of eq.~\eqref{eq:expFaMorris} closest to $0$ correspond to dominant exponential tails of $P(f_a)$. By eq.~\eqref{eq:morrisbarnes1}, we have the following poles:
 \begin{enumerate}
 	\item  $s =- b^{-1}$ from $\Gamma(1 - \tilde{n} b^2)$,
 	\item  $s=-Q $  from $\Barnes_b(Q-\tilde{n}b)$, 
 	\item  $s=2a - Q $  from $\Barnes_b(Q - 2a-\tilde{n}b )$.
 \end{enumerate}
  In the two last cases we used the zero of $\Barnes_b$ in eq. \eqref{eq:barneszeros} with $u=v=0$. Other poles are more negative. Since there are no positive poles, the right tail $P(f_a \to +\infty)$ decays faster than any exponential, which is consistent with the hard wall at $F_a = -Qt$ predicted in the large deviation functions~\eqref{eq:ldf2} and eq.~\eqref{eq:ldf3}.  Concerning the left tail  $P(f_a \to -\infty)$, the competition between the above listed poles results two cases:
\begin{itemize}
	\item[\texttt{i}] $ b/2 < a < Q/2$: the pole $2a-Q$ dominates, giving an exponential left tail $P(f_a) \sim e^{(Q-2a)f_a}$, in agreement with the large deviation function eq. \eqref{eq:ldf2} in the domain $  \hat{y} = F_a/t \in (2 (a-Q),-Q)$. 
	\item[\texttt{ii}] $a < b/2$: the pole $-1/b$ dominates and gives an exponential left tail $P(f_a) \sim e^{f_a/b}$, in agreement with the large deviation function eq.~\eqref{eq:ldf3} in the domain $ \hat{y} \in (-\frac{2}{b} , -Q)$. The other exponential (between LN and B regimes) in eq.~\eqref{eq:ldf3} will be confirmed in section~\ref{sec:morrisLN} below.
\end{itemize}

The above considerations concern the high-temperature phase. For the $\beta > 1$ phase, we can invoke the freezing scenario~\cite{carpentier2001glass,fyodorov2008statistical,fyodorov2009statistical}:
\begin{equation} \left.\overline{e^{s f_a }}\right\vert_{\beta > 1}  \approx e^{s C_\beta}  \frac{\Gamma (1 + s)}{\Gamma(1 + s/\beta)} \, \widetilde{\mathcal{M}}(-s, a, b = 1) \,, \label{eq:freezingfa} \end{equation}
which determines the free energy distribution up to a shift $C_\beta$. Replacing $\widetilde{\mathcal{M}}$ by ${\mathcal{M}}$ results also in a shift of free energy, which is related to the log-correction associated with the freezing transition~\cite{rosso12counting,fyodorov2015high}. The right hand side of eq.~\eqref{eq:freezingfa} has a new pole $s = -1$. In the case \texttt{ii} above, the new pole coincides with $-1 = -1/b$. Therefore, there is a double pole at $s = -1$, and the exponential tail acquires a log-correction:
\begin{equation} P(f_a) \stackrel{f_a \to -\infty} \sim \abs{f_a} e^{f_a} \,,\, \beta > 1 \,,\, a < 1/2 \,. \label{eq:cld} \end{equation} 
We call this a log-correction since $ \abs{f_a} e^{f_a} = e^{f_a + \ln\abs{f_a} }$.
The left tail exhibited in eq.~\eqref{eq:cld} is a universal feature of logREM without charge in the frozen ($\beta>1$) phase~\cite{carpentier2001glass}, and we show here that it prevails for logREM with charge $a < 1/2$. However, when $a > 1/2$, the new pole $-1 < 2a-Q = 2a-2$ [recall $Q=2$ when $\beta>1$, see eq.~\eqref{eq:defQ}] the dominant left exponential tail is still 
\begin{equation} P(f_a) \stackrel{f_a \to -\infty} \sim  e^{2(1-a)f_a} \,,\, \beta > 1 \,,\, 1/2 < a < 1 \,, \label{eq:cldno} \end{equation} 
 without log-corrections. Eqs.~\eqref{eq:cld} and \eqref{eq:cldno} apply both to the zero-temperature limit where $f_a$ becomes the shifted minimum of the logREM with one charge. Compared to logREMs without charge, the universal left tail has a richer behavior, even restricted to the Unbound regime.

When approaching the {Critical} regime from the Unbound regime, $a\nearrow Q/2$ in eq.~\eqref{eq:discont}, the analytically continued Morris integral vanishes [because of a zero of $\Barnes_b(Q-2a)$ in eq.~\eqref{eq:morrisbarnes1}]: 
\begin{equation}{\mathcal{M}} (-s/\beta, \alpha \nearrow Q/2,b) \sim (Q-2\alpha) \,. \label{eq:problemzero} \end{equation}  
Applying this to eq.~\eqref{eq:discont}, we obtain the following asymptotic behavior approaching the Unbound-Critical boundary from the Unbound side:
\begin{equation}
\overline{Z_0^n e^{-a\phi_1} } \sim (Q - 2a) M^{a^2 + Q n \beta} \,,\, a \nearrow Q/2 \,,\, n < 0 \,,
\end{equation}
which agrees with the traveling-wave prediction eq.~\eqref{eq:poleU}.

Eq.~\eqref{eq:problemzero} was already observed in Ref.~\cite{fyodorov2015moments} in the case $s = -n \beta= a$, and was considered a pathology. From the point of view of this work, this pathological zero must be due to the termination point transition from the Unbound to the Critical regime, to which we come now.

\subsection{Critical regime} \label{sec:morriscritical}
In the Critical regime of the $\beta < 1$ phase, the RSB solution becomes $n_0 \beta = Q/2 - a$, $m = 1$, see Table~\ref{tab:res}. Therefore, $n_0$ replicas are attached to $z = -1$, leaving $\tilde{n} = (n-n_0)$ moving charges in the Morris integral, whereas the total charge at $z= -1$ becomes $\alpha = n_0 \beta + a = Q/2$. However, by eq.~\eqref{eq:problemzero}, the Morris integral vanishes exactly at $\alpha  = Q/2$, making a formula analogous to eq.~\eqref{eq:discont} problematic. 

We now employ a trick to provide a heuristic resolution of the problem. For this, it is convenient to rewrite the main observable in the following form:
 \begin{align} 
 & Z_0^{n}  e^{-a \phi_1}  = Z_0^{\ell} e^{-a x}   \\
 &  x \defeq - \beta^{-1}  \ln p_{\beta,1} \geq 0 \,,\, \ell = n  + a / \beta \, \label{eq:defx} \end{align}
where we recall from eq.~\eqref{eq:girnasov} that $p_{\beta,1}$ is the Gibbs probability weight of site $1$. Then, eq.~\eqref{eq:discont} can be rewritten as
\begin{equation}\label{eq:discontgibbs}
\overline{Z_0^{\ell} e^{-a x}}  \approx e^{t(a^2 - Qa + Q \ell \beta)} \mathcal{M}( \ell - a/\beta, a, \beta) 
\end{equation}
in the Unbound regime, where $t=\ln M$. The left hand side, as a function of $a$, is the Laplace-Fourier transform of $P_\ell(x') := \overline{Z_0^{\ell} \delta(x - x')}$. By eq.~\eqref{eq:defx}, $P_\ell(x') = 0$ for $x' < 0$. Therefore, denoting by $ \theta(x')$ the Heaviside function, we have $P_\ell(x') = P_\ell(x') \theta (x') $. Taking the Laplace-Fourier transform of this equation, we turn the product into a convolution, obtaining:
\begin{align}\label{eq:criticalmagic}
&\overline{Z_0^{n} e^{-a \phi_1}}  \sim  \int_{\mathcal{C}} \frac{\dif \alpha}{2\pi \im }  e^{t(\alpha^2 - Q\alpha + Q (n\beta + a))} C_{\mathcal{M}}(\alpha)   \,,\\
&  C_{\mathcal{M}}(\alpha) \defeq \frac{\mathcal{M}( n + (a - \alpha)/\beta, \alpha, \beta)}{a - \alpha} \label{eq:CMalpha}
\end{align}
where the contour in vertical and satisfies $\Re(\alpha) < a$. The denominator comes from the Laplace-Fourier transform of the Heaviside function. Now eqs.~\eqref{eq:criticalmagic} and \eqref{eq:CMalpha} make sense in both Unbound and Continuum regimes, and do not suffer from the ``problematic zero'' of the Morris integral. So we propose eq.~\eqref{eq:criticalmagic} as an extension of eq.~\eqref{eq:discontgibbs} into the Critical regime.  

Notice that eq.~\eqref{eq:criticalmagic} is similar to eq.~\eqref{eq:HS2} in the traveling-wave approach and can be analyzed by a saddle-point approximation. In particular, the exponent in $ e^{t(\alpha^2 - Q\alpha + Q (n\beta + a))}$ is identical to that in eq.~\eqref{eq:HS2}, so the saddle point is $\alpha = Q/2 $. In the Unbound regime, using eq.~\eqref{eq:morrisbarnes1}, we can check that the pole $\alpha  = a$ is first crossed to move the contour towards the saddle point. The discrete term is just the right hand side of eq.~\eqref{eq:discont}. Thus, eq.~\eqref{eq:criticalmagic} recovers the Unbound regime result.

Now, in the Critical regime, $a > Q/2$, $a + n\beta < Q/2$,  no pole is crossed to move the contour to cross the saddle point, where eq.~\eqref{eq:criticalmagic} has a simple zero, by eq.~\eqref{eq:problemzero}. As a consequence, we obtain the correct leading behavior and log-correction predicted by the traveling-wave and LFT approach:
\begin{align}
\overline{Z_0^{n} e^{-a \phi_1}}  \sim & \,  e^{t (Q(n\beta+a) -Q^2/4)} \, t^{-\frac32} \left(-C_{\mathcal{M}}''(Q/2) \right) \label{eq:32morris}   \,. 
\end{align}
We can further determine the divergence of the amplitude when approaching the U/C or U/B phase boundary. For this, we retain the relevant zero and poles of $C_{\mathcal{M}}(\alpha)$ as follows:
\begin{equation}
C_{\mathcal{M}}(\alpha \to Q/2) \sim \frac{(Q-2\alpha)}{(Q - a - \alpha - n \beta) (a-\alpha)} \,. \label{eq:CMasym}
\end{equation}
where the first pole is due to a zero of $\Barnes_b(Q - 2\alpha - \tilde{n}b)$ (with $\tilde{n}b = a+ n\beta - \alpha$) in the denominator of eq.~\eqref{eq:morrisbarnes1}. Then 
\begin{equation} C_{\mathcal{M}}''(Q/2) \sim \frac{n \beta}{\left(a-\frac{Q}{2}\right)^{2} \left(a+\beta  n-\frac{Q}{2}\right)^2 } \,. \end{equation}
These divergences are in agreement with those of the traveling-wave prediction, eq.~\eqref{eq:32}.
 On the U/C (or U/B) phase boundary, $C_{\mathcal{M}}(Q/2) \neq 0$ since a pole cancels the zero in eq.~\eqref{eq:CMasym}, then the saddle point approximation yields a $t^{-1/2}$ log-correction, in agreement with the LFT and traveling-wave approach. 

The proposal eq.~\eqref{eq:criticalmagic} is a heuristic suggestion indicating how the replica approach to integrable logREMs can be adapted to a regime or phase with broken replica symmetry. It connects nicely the treatment of the replica-symmetric (Unbound, high-temperature) case and the discrete-continuum aspect of the LFT and traveling-wave equation approaches of the main text. It provides also an explanation of the following phenomenon: the ``problematic zero'' of the Coulomb gas integral is responsible for the log-correction with exponent $3/2$ in the Critical regime, beyond the termination point transition~\cite{cao17fbm}. Note that the same phenomenon was also observed in the frozen ($\beta > 1$) phase, and an apparently different explanation was given in Refs.~\cite{fyodorov2015high,rosso12counting}. It will be interesting to clarify the relation between them. 

The above being said, we believe that eq.~\eqref{eq:criticalmagic} does \textit{not} contain all the model-dependent corrections in the Critical Regime. Leaving a complete treatment to further study, we point out one important correction that is missing. As argued in Ref.~\cite{cao16order}, when replicas form groups of size $>1$, the internal structure of the group is non-trivial and contributes a factor depending on the short-distance details (``UV-data'' )  of the model, in addition to the Coulomb gas integral, which depends only on the long-distance details (``IR-data''). The short-distance factor is known to have tangible effects, e.g., on the distribution of the second minimum of logREMs~\cite{cao16order}, but its analytical behavior is poorly understood. Now the right hand side eq.~\eqref{eq:criticalmagic} does not contain such a short-distance factor, $C_{\text{UV}}(\alpha)$.  However, we know~\cite{cao16order} that this factor becomes trivial when the replica symmetry is unbroken, i.e., in the Unbound regime (of the $\beta<1$ phase): $C_{\text{UV}}( \alpha \to a)  \to 1$, so we expect that the asymptotic behaviors above are not affected by its presence, at least not too far away from the U/C boundary.

\subsection{Bound regime} \label{sec:morrisbound}
In the \textit{Bound} regime, all the replicas are attached to the charge in the RSB solution. Therefore,  the corresponding Morris integral will have no moving charges: $\tilde{n} = 0$, $\alpha =  a + n\beta$, so that the analogue of eq.~\eqref{eq:discont} becomes trivial in the Bound regime (of the $\beta < 1$ phase):
\begin{equation}
 \overline{Z_0^n e^{-a\phi_1} } M^{\Delta(a,n)}  \sim \mathcal{M}(0, a + n \beta, \beta) = 1 \label{eq:discontbound}
\end{equation}
where $\Delta(a,n) = -(a + n\beta)^2 $. Although the leading behavior is correct, we cannot extract further corrections from integrability. Such corrections can nonetheless come from a short-distance factor describing the replica group attached to the charge at $z = -1$. 

In a related note, let us comment on the 2d logREM considered in Ref.~\cite{cao16liouville} and Ref.~\cite{cao17fbm}, Section~III B. Its random potential is the sum of 2d GFF on the complex plane (with a short-distance cut-off $\epsilon$ and a large-distance cut-off $L$) and a deterministic background potential with two charges: $U(z) =  4 a_1 \abs{z / L} + 4 a_2 \abs{(z-1) / L}$, such that $a_1, a_2 < Q/2$ and $a_1 + a_2 > Q/2$.  In the thermodynamic limit $\epsilon \to 0, L \to \infty$, the model has a delicate behavior: it is in the Unbound phase with respect to each of the charges at $z = 0$ and $z = 1$, but is in the Bound phase with respect to the background potential coarse-grained to the scale $L$: $U(z) \approx 4 a \abs {z/L}$, $a  = a_1 + a_2$. We shall focus on the latter point of view. Then, the results of the main text on logREMs with one charge apply:
the observable $\overline{Z_a^n} =\overline{\exp(s F_a)}$ ($s = -n \beta$) is governed by the Bound regime if $a  + n \beta > Q/2 \Leftrightarrow s < a - Q/2 $ and by the Critical regime otherwise. Now, in the Bound regime, the model-dependent corrections of  $\overline{\exp(s F_a)}$ can be calculated by an exactly solvable 2d Coulomb gas integral, the Dotsenko-Fateev (DF) integral~\cite{dotsenko1984conformal} (see also~\cite{cao17fbm}, eq. 37):
\begin{equation}\label{eq:DF}
\overline{\exp(s F_a)} \sim \epsilon^{2Qs} M^{-2as + s^2}  \mathcal{D}(s\vert a_1, a_2) \,,\, 
\end{equation}
where $M=L^2$ and $\mathcal{D}(s\vert a_1, a_2)$  is the analytically continued DF integral (it is in fact equal to $ C^{\text{DOZZ}} (a_1, a_2, Q-a_1-a_2 + s)s / \Gamma(1 + s/\beta)$, in terms of the DOZZ structure constant of LFT~\cite{zamolodchikov1996conformal}, see also~\cite{cao17fbm}). A key property of $\mathcal{D}(s\vert a_1, a_2)$ is that it vanishes when $s$ approaches Bound/Critical boundary:
\begin{equation}
\mathcal{D}(s\vert a_1, a_2) \sim a - Q/2 - s \,,\, s \nearrow a - Q/2 \,. \label{eq:DFzero}
\end{equation}
This zero would seem puzzling as it naively implies that the moment generating function in the left hand side of eq.~\eqref{eq:DF} vanishes at $s = a-Q/2$, which is impossible. However, by now, we are used to this ``problematic zero'': it is not pathological, but an expected signature of the transition from the Bound regime without log-correction to the Critical regime with log-correction. We remark that there is a striking similarity between this phenomenon and the vanishing of the continued Coulomb gas integral, eq.~\eqref{eq:problemzero}, which signals the Unbound$\to$Critical transition. This suggests that the U/C and B/C transitions have the same nature; such an observation was also made in Section~\ref{sec:LDF1B}, from a large deviation theory perspective. From the LFT point of view, this is rather expected, and we may further speculate that the U and B regimes can be mapped to each other by a conformal mapping such as $z \mapsto 1/z$, which maps short and large distances.


\subsection{Log-Normal regime}\label{sec:morrisLN}
In the \textit{Log-normal} regime, the RSB solution is $n_0 = 0$ and $m = n$, see Table~\ref{tab:res}. All the $n$ replicas form a same group, but are not attached to the charge. Therefore, the continuum limit of the replica sum eq.~\eqref{eq:wick} corresponds to the Morris integral with only one moving charge of $b = n \beta$:
\begin{align}
\overline{Z_0^n e^{-a \phi_1}} & \sim M^{1 + n^2\beta^2 + a^2}  \frac{\Gamma (1-2 a n \beta )}{\Gamma (1-a n \beta )^2}  \,.  \label{eq:morrisLN0}
\end{align}
Applying the Girsanov transform eq.~\eqref{eq:girnasov} to eq.~\eqref{eq:morrisLN0} we obtain:
\begin{equation}
\overline{\exp(s F_a)} \, \approx  M^{1 + s^2} \frac{\Gamma (1 + 2 a s )}{\Gamma (1 + a s )^2}  \label{eq:morrisLN}
\end{equation}
Now, eq.~\eqref{eq:morrisLN} has its least negative pole at $s = -1/(2a)$, which corresponds to an exponential left tail $P(F_a) \sim e^{F_a / 2a}$. The exponent  $1/(2a)$ agrees with the large deviation prediction eq.~\eqref{eq:ldf3}, in the domain $-\frac1a - 2 a <  \hat{y}  < -\frac1a$.
Combined with the results of Section~\ref{sec:morrisU}, we recovered all three exponential tails predicted by the large deviation function~eq. \eqref{eq:Lf}. We recall that the latter was obtained under the convexity assumption. These results constitute a non-trivial test of this hypothesis.


\section{Two-variable large deviation function: Log-Normal and interpolation regime}\label{sec:fxymore}
\begin{figure}
	\includegraphics[width=\columnwidth]{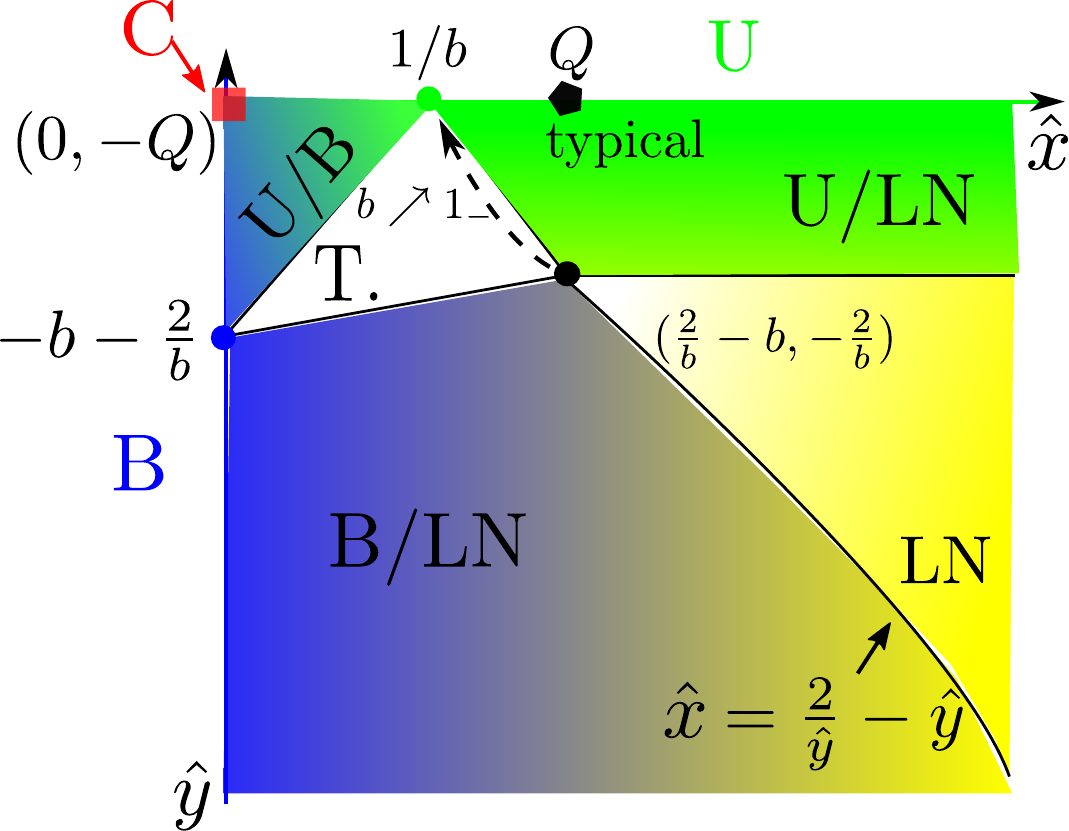}
	\caption{The regimes of the generalized multi-fractal spectrum $f(\hat{x}, \hat{y})$ of logREM without charge, defined in eq. \eqref{eq:multifracdef}. The expressions for U,C and B regimes are given in equations \eqref{eq:multifrac}. The other first-order interpolation regimes are given in eqs.~\ref{eq:fxymore}. The whole triangular region corresponds via Legendre transform to the triple point $\mathrm{T}$ described in Fig.~\ref{fig:phase}. Note that the Unbound and Bound regimes occupy the whole semi-axes $\hat{y} = -Q$ and $\hat{x} = 0$, respectively. The noted points on them are not critical.  When $\beta \geq 1$, both the T and U/LN interpolation regions become degenerate.} \label{fig:phase2}
\end{figure}

We provide here the explicit expressions of the two-variable large deviation function $f(\hat{x}, \hat{y})$, defined in eq.~\eqref{eq:multifracdef}, in the interior of the parameter plane complementing eqs.~\eqref{eq:multifrac}. The results below are summarized in Fig~\ref{fig:phase2}. We recall from eq.~\eqref{eq:legendre1} that $f(\hat{x}, \hat{y})$ and $\Delta(a,n)$ (in Table~\ref{tab:res}) are Legendre transform of each other, under the following duality of variables: 
\begin{equation} a = \partial_{\hat{x}} f \,,\, a + n \beta = \partial_{\hat{y}}  f \,. \label{eq:legendredual} \end{equation}
We begin by calculating $f(\hat{x}, \hat{y})$ in the Log-Normal regime by Legendre transforming $\Delta = -a^2 - (n\beta)^2 -1$:
\begin{subequations}\label{eq:fxymore}
\begin{align}
&f(\hat{x}, \hat{y}) = - \frac{\hat{x} ^2}{4} - \frac{\hat{x}  \hat{y}}{2} - \frac{\hat{y}^2}2 + 1   \nonumber \\
&\hat{y} < -2b^{-1} \,,\, \hat{x} >  2 \hat{y}^{-1} - \hat{y} \label{eq:multifracLN}
\end{align} 
The regime boundary in the $(\hat{x}, \hat{y})$-plane is obtained by apply the Legendre transform $\hat{x} = -2a + 2 n \beta, \hat{y} = -2 n \beta$ to the Log-normal region $ n \beta \in (1/b, 1/(2a)) $, see Table~\ref{tab:res}.

The value of $f$ in the rest of the $(\hat{x}, \hat{y})$-plane can be calculated by convexity interpolation. This results in four regions, corresponding to the three first-order transitions and their joining point $(a = b/2, n\beta = b^{-1})$, respectively:
\begin{itemize}
\item[-] U/LN boundary:
\begin{align}
&f(\hat{x}, \hat{y}) = \frac{\hat{y} + Q}{b} - \frac14 (\hat{x} + \hat{y})^2 \, \nonumber  \\ 
&-2/b < \hat{y} < -Q \,,\, \hat{x} > -  b+\hat{y}  
\end{align} 
One can check that $f(\hat{x}, \hat{y})$ connects continuously eq.~\eqref{eq:multifracU} and \eqref{eq:multifracLN}, with continuous first derivatives for the latter, and satisfies $  \partial_{\hat{y}}  f - \partial_{\hat{x}} f = 1/b $, which corresponds to the U/LN boundary $(a+n \beta) - a = 1/b$ in Fig.~\ref{fig:phase} by the Legendre duality eq.~\eqref{eq:legendredual}.  
\item[-] B/U boundary: 
\begin{align}
& f(\hat{x}, \hat{y}) = Q \hat{x}-\frac14 (\hat{y}-\hat{x})^2 ,\,  
\hat{x} + b- \frac2b < \hat{y} <  -Q \,. 
\end{align}
One can check that $f$ connects continuously eq.~\eqref{eq:multifracU} and eq.~\eqref{eq:multifracB}, and satisfies $\partial_x f + \partial_y f= Q$, which corresponds to the B/U boundary $a + (a  +  n \beta) = Q$ in Fig.~\ref{fig:phase} by the Legendre duality eq.~\eqref{eq:legendredual}. The boundary $\hat{y} = \hat{x} + b- \frac2b$ corresponds to the Triple point $\partial_{\hat{x}} f = a = b/2, \partial_{\hat{y}} = a + n \beta = b/2 + 1/b$ in Fig.~\ref{fig:phase}.
\item[-] Triple point:
\begin{align}
& f(\hat{x}, \hat{y}) = \frac{b\hat{x}}{2}  + \left(\frac1b + \frac{b}2\right) \hat{y}  +\left(\frac1b + \frac{b}2\right) ^2
\end{align}
in the triangle spanned by $(0, -2/b - b), (1/b, -Q)$ and $( 2/b-b ,-2/b)$, as depicted in Fig.~\ref{fig:phase2}. The spectrum is linear with $\partial_{\hat{x}} f = a = b/2, \partial_{\hat{y}} f = a + n \beta = b/2 + 1/b$, and connects continuously to the neighboring regions.
\item[-] B/LN boundary: 
\begin{align}
&f\left( \left(\frac1a- 2a \right) u, 2a (u - 1) - \frac1a \right)  \nonumber \\
=& -\frac{1}{4}\left(2a + \frac1a \right)^2 +  2 u \,. \nonumber \\
& a > b/2 \,,\, 0 < u < 1 \,.  \label{eq:BLN}
\end{align} 
The first line is obtained as the solution to the non-linear first order PDE $ \partial_{\hat{x}} f(\partial_{\hat{y}} f -  \partial_{\hat{x}} f) = 1/2$ with the boundary condition set by matching with the B regime eq.~\eqref{eq:multifracB}. We obtained it by applying the standard characteristic method for non-linear first order PDE's. Indeed, $u$ parametrizes of the characteristic curves. $u = 0$ connects to the B regime ($\hat{x}=0$) and $u = 1$ connects to a boundary of LN regime $\hat{x} = 2/\hat{y} - \hat{y}$ (see Fig.~\ref{fig:phase2}), across which $f$ has continuous first derivatives.  We haven't found an explicit formula which is more compact than the parametric form eq.~\eqref{eq:BLN}.
\end{itemize}
\end{subequations}
Note that both the triple point and the U/LN boundary regions degenerate in the $\beta \geq 1$ phase, where the latter two points of the triangle collapse.

\section{Derivation of traveling-wave equations}\label{sec:kppderivation}
In this appendix, we outline the derivation of eqs.~\eqref{eq:kppH} and \eqref{eq:kppinitI}. As a warm-up, it is helpful to recall the derivation of the classic KPP equation~\eqref{eq:kpp} in a broader generality~\cite{makean75bbm,derrida1988polymers}. 
For this, let $\phi_j(t), j = 1, \dots, M(t)$ be the particle positions of a BBM at time $t$, and $\Theta(y)$ be any well-behaved function defined on the real line. Then, it is well-known that the observable 
\begin{equation} G(y,t) \defeq  \overline{\prod_{j=1}^{M(t)}\Theta(y-\phi_j(t))} \label{eq:defGgen}  \end{equation}
satisfies the KPP equation with initial condition
\begin{align}
G_t = G_{yy} + (G-1)G \,,\, G(y,0) = \Theta(y) \,. \label{eq:kppgen}
\end{align}
In particular, by taking $\Theta(y) = \exp(-e^{\beta y})$, we obtain eq. \eqref{eq:kpp}. Eq. \eqref{eq:kppgen} is derived as a backward master equation, by considering what happens in $t \in [0, \dif t]$ to the initial particle. Its diffusion leads to the term $G_{yy}$. With probability $\dif t$ It splits into 2 particles, one offspring is labelled $1$ and the other $2$. Then we may bookkeep the change of $G$ as:
\begin{align}
G(y) &= \overline{\prod_{j} \Theta_j } \leadsto  
\overline{ \prod_{\text{(1)}} \Theta_j} \, \overline{ \prod_{\text{(2)}} \Theta_j }
= G(y) G(y) \label{eq:nonlinear}
\end{align}
where $\Theta_j := \Theta(y-\phi_j(t))$, and $\prod_{(1)}$ means the product over all offsprings of $1$, and similarly for $2$. The difference $(G^2 - G)$ gives the non-linear term in eq. \eqref{eq:kppgen}.

After the above warm-up, we now come to eq. \eqref{eq:kppH}. Again, more generally, we show that any observable of type 
\begin{equation}  H(y,t) \defeq \overline{\Xi(y - \phi_1(t)) \prod_{j=2}^{M(t)} \Theta(y-\phi_j(t))} \,, \label{eq:defHgen}  \end{equation}
where $\Xi$ is any well-behaved function defined on the real line, satisfies the KPP-type equation
\begin{equation}
H_t = H_{yy} + (G-1)H \,,\, H(y,0) = \Xi(y) \,. \label{eq:kppHgen}
\end{equation}
where $G$ is the solution of eq. \eqref{eq:kppgen}. In particular, eq. \eqref{eq:kppH} is obtained by $\Theta(y) = \exp(-e^{\beta y})$ and $\Xi(y) = e^{a y} \Theta(y)$. Eq. \eqref{eq:kppHgen} is also derived as a backward equation as eq. \eqref{eq:kppgen}. The only difference is the non-linear term $G H - H$, which can be given by a similar consideration as eq. \eqref{eq:nonlinear}:
\begin{align*}
H(y) &= \overline{ \Xi_1 \prod_{j\geq 2} \Theta_j } \leadsto \overline{ \Xi_1 \prod_{\text{(1)}} \Theta_j} \, \overline{ \prod_{\text{(2)}}\Theta_j} 
= H(y) G(y) \,,
\end{align*}
where $\Xi_1 :=\Xi(y-\phi_1(t)).$

We now derive eq.~\eqref{eq:kppI} used in the common length distribution computation. For this we consider again the splitting of the initial particle into $1$ and $2$ during the  time interval $[0, \dif t]$ where $\dif t < \hoverlap$. Then, both of the two particles $j,k$ in eq.~\eqref{eq:defI} are offsprings of either $1$ or $2$,  so that the analogue of eq.~\eqref{eq:nonlinear} is (see Fig.~\ref{fig:overlap}):
\begin{align*}
I(y,\hoverlap,\tau) \leadsto I(y,\hoverlap, \tau) G(y,\hoverlap + \tau) +G(y,\hoverlap + \tau)  I(y,\hoverlap, \tau)   \,,
\end{align*}
resulting in the non-linear term $(2G(y,\hoverlap + \tau) - 1) I $ in eq.~\eqref{eq:kppI}. The main difference from eq.~\eqref{eq:kppHgen} is that the ``marked'' particle (here the common ancestor of $j$ and $k$) is summed over rather than fixed, hence the extra factor $2$. 

To obtain the initial condition eq.~\eqref{eq:initI}, we compute directly $I(y, \hoverlap = 0, \tau)$ by the definition eq.~\eqref{eq:defI}. $\hoverlap = 0$ implies that the initial particle splits initially into two particles $1$ and $2$. The sum over $j,k$ in eq.~\eqref{eq:defI} is reduced, by the requirement $\hoverlap_{jk} = 0$, to a sum of $j$ over the offspring of $1$, and of $k$ over the offspring of $2$ (both at time $\tau$), or vice versa. It is then not hard to see that $I(y,\hoverlap = 0,\tau) $ is $2$ times a product of two identical factors:
\begin{align*} 
I(y,\hoverlap = 0,\tau) &= 2 J(y,\tau)^2 \,,\, \text{where} \\
J(y,\tau) =& \sum_{k} \overline{e^{\beta (y-\phi_k (\tau))} \prod_{j} \exp(-e^{\beta(y-\phi_j(\tau))})}
\\ =& -\beta^{-1} \partial_y \overline{\left[ \prod_{j} \exp(-e^{\beta(y-\phi_j(\tau))}) \right]} \\ 
=& - \beta^{-1} G'(y, \tau) \,,
 \end{align*}
 where $G$ is the solution to eq.~\eqref{eq:kpp}. Thus we obtained the initial condition eq.~\eqref{eq:initI}.

\bibliography{rems.bib}

\end{document}